\documentclass[preprint, 12pt]{elsarticle}
\usepackage{bm, color, amssymb, amsmath,
	booktabs,siunitx,subfigure,mathtools,natbib,hyperref}
\usepackage[version=4]{mhchem}
\biboptions{numbers,sort&compress}

\usepackage[displaymath, mathlines,running]{lineno}

\newcommand{\bi}{\bm}

\journal{Journal of Power Sources}

\usepackage[utf8]{inputenc}
\usepackage{etoolbox}

\begin{document}

\begin{frontmatter}
\title{Thermally driven convection in Li$||$Bi liquid metal batteries}

\author[hzdr,poliTo]{Paolo Personnettaz}
\author[hzdr]{Pascal Beckstein}
\author[hzdr]{Steffen Landgraf}
\author[hzdr,usa]{Thomas K\"ollner}
\author[hzdr]{Michael Nimtz}
\author[hzdr]{Norbert Weber}
\author[hzdr]{Tom Weier}

\address[hzdr]{Helmholtz-Zentrum Dresden -- Rossendorf, %
Bautzner Landstr.\ 400, 01328 Dresden, Germany}

\address[poliTo]{Politecnico di Torino, Corso Duca degli Abruzzi 24, 10129
Torino, Italy}

\address[usa]{University of California, Santa Barbara, CA 93106, USA}

\begin{abstract}
  Liquid Metal Batteries (LMBs) are a promising concept for cheap electrical
energy storage at grid level. These are built as a stable density stratification
of three liquid layers, with two liquid metals separated by a molten salt. In
order to ensure a safe and efficient operation, the understanding of transport
phenomena in LMBs is essential.  With this motivation we study thermal
convection induced by internal heat generation.
We consider the electrochemical nature of the cell in order to define the heat
balance and the operating parameters. Moreover we develop a simple 1D heat
conduction model as well as a fully 3D thermo-fluid dynamics model. The latter
is implemented in the CFD library OpenFOAM, extending the volume of fluid
solver, and validated against a pseudo-spectral code. Both models are used to
study a rectangular 10$\times$10\,cm Li$||$Bi LMB cell at three different states
of charge.
\end{abstract}

\begin{keyword}
liquid metal batteries \sep heat transfer \sep thermal convection \sep
thermodynamics \sep OpenFOAM \sep volume of fluid \sep spurious currents 
\end{keyword}

\end{frontmatter}
\clearpage

%\linenumbers

\section{Introduction}
Grid-level storage will be an indispensable ingredient of future
energy systems dominated by volatile renewable electricity sources
\cite{Lindley:2010}. As of today, storage options are limited and far
from available in the required capacity \cite{Pickard:2015}. Liquid
metal batteries (LMBs) might help to economically bridge the storage
gap \cite{Spatocco2015}.

\begin{figure}[bht!]
\centering
\includegraphics[width=0.8\textwidth]{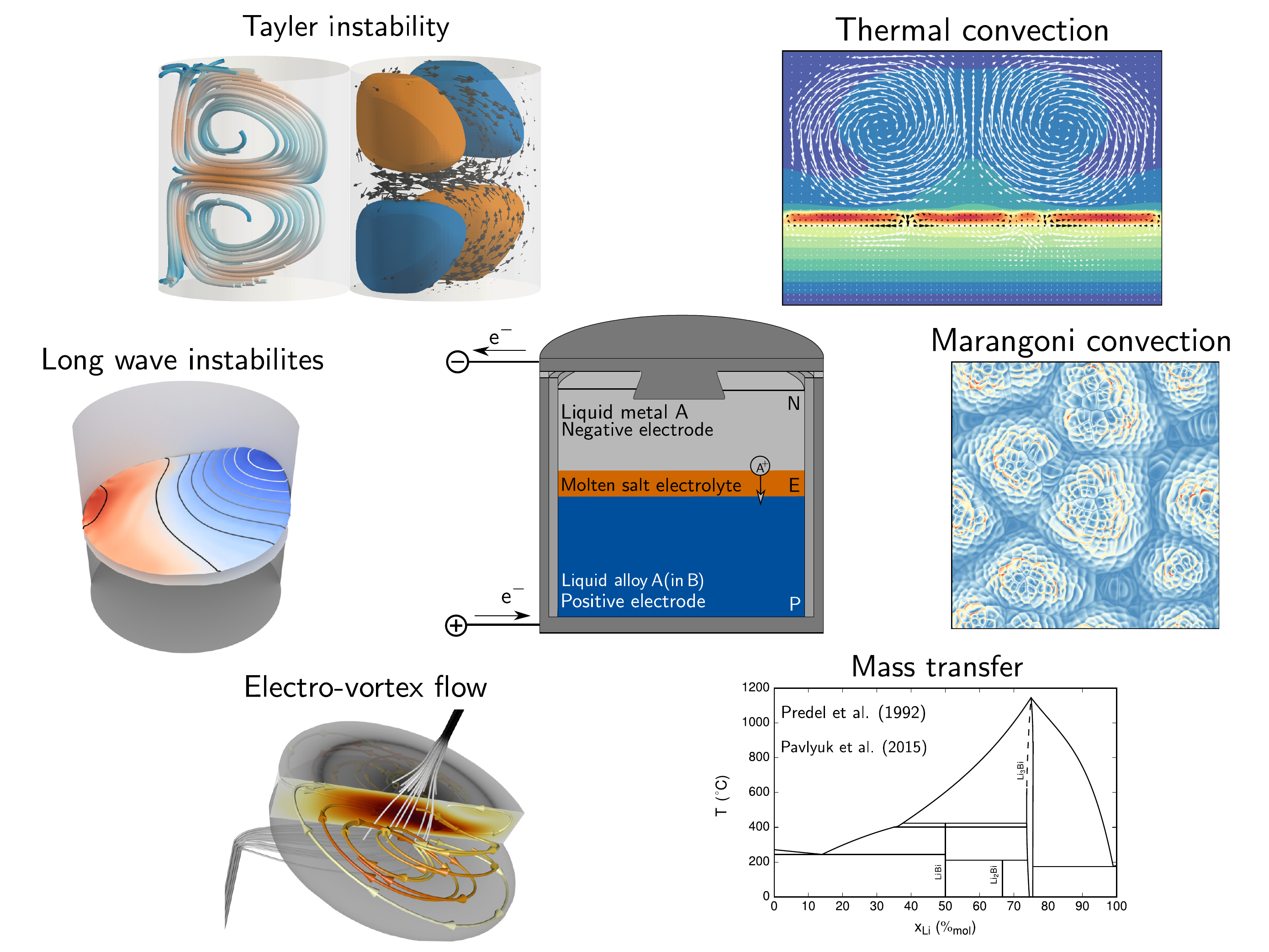}
\caption{Sketch of an LMB with typical inventory (middle) and expected
fluid dynamics and transport phenomena in LMBs.}
\label{f:flow}
\end{figure}

An LMB consists of a low-density liquid metal negative electrode, an
intermediate-density molten salt electrolyte, and a high-density
liquid metal positive electrode (Fig.~\ref{f:flow})
\cite{Kim2013b}. As the name indicates, the operating temperature is
such that each phase is in a liquid state.  Typical negative electrode
materials are K, Li, Na \cite{Agruss1963a,Cairns1967,Cairns1969b} and more
recently Ca \cite{Kim2013a} and Mg \cite{Kim2013b}. Positive
electrodes can consist of Bi, Pb, Sn \cite{Cairns1967}, Hg
\cite{Agruss1963a}, Se, Te \cite{Cairns1969b}, as well as Sb
\cite{Kim2013b} and other metals. 
From the perspective of fluid dynamics, LMBs are multi-physics systems
that couple electrochemical reactions, thermal effects, mass
transport, electric currents, magnetic fields, and phase changes with
fluid flow. In the last years, several studies have been dedicated to
model time and space resolved physical processes in LMBs, for a recent
review see \cite{Kelley2017}. These studies were aimed at uncovering the
non-equilibrium processes during operation as well as at addressing
crucial design problems.  A central issue has been the mechanical
integrity of the electrolyte layer. Its height is governed by
conflicting objectives, i.e., the reduction of ohmic losses (thin
electrolyte) versus safety against short circuits (thick electrolyte).

Fluid-dynamical research has focused so far mainly on 'differential
density cells' containing three liquid layers in stable stratification
but free to move (Fig.~\ref{f:flow}, center). Electrochemical cell
characteristics are often reported for configurations featuring a
retainer for the negative electrode, typically a metal foam, e.g.,
\cite{Ning2015}.

Several studies have shown that fluid motion can be triggered by
magneto-hydrodynamic effects, i.e., the interaction of a magnetic
field (either a background field or the field caused by the battery
current) with the cell current. This includes the Tayler instability
\cite{Stefani2011,Weber2013,Weber2014,Herreman2015,Weber2015b,Weier2017},
electro-vortex flows
\cite{Bradwell2015,Weber2014b,Stefani2015,Ashour2017a,Weber2018} as
well as interface instabilities
\cite{Zikanov2015,Weber2017,Weber2017a,Bojarevics2017,Horstmann2017,Zikanov2017}.
While the Tayler instability will get substantial only for large
system (in the order of meters) \cite{Herreman2015}, electro-vortex
flow will appear already in small cells \cite{Stefani2015}. Long wave
interface instabilities may endanger the safe operation of medium to
large size LMBs \cite{Weber2017a,Horstmann2017}.

For laboratory scale experiments, thermal convection is an important
source of fluid motion. Temperature gradients (producing
forces by density and interfacial-tension gradients) seem to be
unavoidable due to either heating or cooling from the surroundings.
In this context, Wang et al.~\cite{Wang2015} studied pure heat conduction in a
single cell. Kelley et al.~\cite{Kelley2014,Perez2015}
experimentally showed how buoyant
convection can be enhanced by applying a current to a liquid-metal electrode
heated from below \cite{Beltran2016}. Shen and Zikanov \cite{Shen2015}
studied numerically the buoyant convection driven by ohmic heating in
a three-layer liquid metal battery with plane, horizontal interfaces. 
This work suggests that heating always triggers convection in the electrolyte
layer for practical configurations. Shen and Zikanov \cite{Shen2015}
further found the influence of the discharge current's magnetic field
on thermal convention to be negligible for laboratory scales.
K\"ollner et al.~\cite{Koellner2017} extended the work of
Shen and Zikanov \cite{Shen2015} by taking into account the
temperature dependency of the interfacial tensions and the different
transport coefficients of the layers.
Choosing a typical material combination of Li separated by
\ce{LiCl-KCl}
from \ce{Pb-Bi}, K\"ollner et al.~\cite{Koellner2017} first analyzed
the linear stability of the stationary pure conduction
state. Furthermore, they performed three-dimensional direct-numerical
simulations using a
pseudo-spectral method while varying electrolyte layer heights, cell
heights, and current densities. 
Four instability mechanisms were identified: 1) buoyant convection in
the upper electrode, 2) buoyant convection in the 
molten salt layer, and 3,4) Marangoni convection at both interfaces between molten 
salt and electrode. The instability mechanisms are partly coupled to each 
other. K\"ollner et al.~\cite{Koellner2017} could confirm the Shen and
Zikanov's \cite{Shen2015} conclusion that buoyant convection
in the salt layer is the most critical one. However, this convection might be
suppressed for small layer heights and low current densities. The Marangoni
effect was found to support buoyant convection as long as the electrolyte
layer thickness remains below one tenth of the upper electrode
height. For thicker Li electrodes, buoyant convection is most active
in the upper layer causing interfacial stresses to counteract convection.  

Recent progress notwithstanding, modeling of LMBs is far from being
comprehensive. Temperature variations discussed so far in the
literature were based only on a few selected terms of the actual
internal energy balance as given, e.g., by \cite{Newman2004}. One of
the so far disregarded terms is the electrochemical heat,
which can be estimated using equilibrium thermodynamics.

This heat contribution warms or cools the positive electrode whenever
the ions of the negative electrode metal alloys or dealloys with the
positive electrode material. A similar effect was found in simple
ternary system of organic solvents with interfacial reactions
\cite{Heines1972}. Experiments that involved exothermic interfacial
reaction revealed intriguing fluid motion \citep{Eckert1999}.

The paper at hand will extend the previous work of Shen and
Zikanov~\citep{Shen2015} and K\"ollner et al.~\citep{Koellner2017} by 
1) including the electrochemical heat, 2) allowing the interfaces to 
be deformable, 3) considering different states of charge, and 4) 
including the lateral walls.
The article proceeds as follows: in section \ref{s:thermodynamics} we
discuss the electrochemistry of the cell and we derive the electrochemical heat. Section \ref{s:thermalPhenomena} gives a
discussion of possible thermal phenomena relevant for an LMB. We
derive a model for the pure conduction state and we describe the
simulation model. Finally, the results are discussed and summarized in
a conclusion.

\section{Thermodynamics and electrochemistry}\label{s:thermodynamics}
The characterization and quantification of the heat sources during
cell charge and discharge is later used in the modeling and simulation
of the thermal processes in an LMB cell. 
This requires the study of the thermodynamic and electrochemical processes in
such a cell.\\ 
The chosen chemistry in this study is \ce{Li||Bi}, which was
experimentally investigated from the electrochemical perspective
by Ning et al. \cite{Ning2015}. Furthermore, there is 
extensive literature (compared to other electrode  pairs) on the thermodynamic
properties of \ce{Bi-Li}
alloys \cite{Grube1934,Zintl1935,Seith1937,Shchukarev1957,
Hansen1958,Foster1964,Demidov1973,
Saboungi1978,Predel1979,Gasior1994,Liu2013,Cao2014}, allowing for good
estimations. The considered molten salt electrolyte is \ce{KCl-LiCl}, with a
near eutectic composition (\SI{41.5}{\percent}$_\text{mol}$ \ce{KCl}). It is not
stable in the cell environment \cite{Foster1964}, but it is the only one for
which all material properties are available.\\
The cell type studied is a fully liquid binary concentration cell.
In the positive electrode, during discharge, oxidized lithium ions and 
bismuth atoms form a fully liquid single phase alloy (\ce{Li (in Bi)} or $\ce{Li_{(\ce{Bi})}}$)
\cite{Ning2015}. This condition is satisfied up to a certain molar fraction of
lithium (liquidus line), $x_\text{liq}(T)=$ \SI{39.5}{\percent}$_\text{mol}$, at
the temperature of interest $T=\SI{450}{\celsius}$. 
For a higher concentration of lithium, the solid intermetallic phase \ce{Li3Bi}
starts to be deposited \cite{Ning2015}. This second
region of operation is not investigated, due to the fact that the
presence of a floating solid phase introduces further complexities
from a thermodynamic and thermo-fluid dynamic point of view. \\
The cell is studied at three different molar fractions of lithium in bismuth
$x_\text{Li}$, at 1, 10 and \SI{38}{\percent}$_\text{mol}$ \ce{Li_{(Bi)}}.

\subsection{Equilibrium cell quantities}\label{s:equilibriumquantities}
The equilibrium cell voltage and the electrochemical heat are 
derived in this paragraph. Both quantities are needed later to compute
the polarization curves, and to simulate thermal phenomena in the LMB.

The equilibrium cell voltages can be taken from the experimental results of
electrochemical studies of \ce{Bi-Li} alloys
\cite{Foster1964,Demidov1973,Gasior1994}.
It is directly related to the variation of partial
molar Gibbs free energy of Li in Bi $\Delta\overline{G}_\text{Li}$ with respect
to the pure Li state.
The equilibrium cell voltage $E_\text{cell,eq}$ is 
\begin{linenomath*}
\begin{equation}
E_\text{cell,eq} =
-\frac{\Delta\overline{G}_\text{Li}}{n_{\text{e}^\text{-}}\cdot\text{F}}\ ,
\end{equation}
\end{linenomath*}
in which  $n_{\text{e}^\text{-}}$ is the number of electrons
transferred per ion ($n_{\text{e}^\text{-}}=1$ for \ce{Li||Bi}) and $\text{F}$ is
the Faraday constant \cite{Kim2013b}. \\
The electrochemical heat $\dot{Q}_\text{r}$ can be derived from the
first and second principle of thermodynamics applied to the cell, assuming a
uniform concentration of \ce{Li} in the positive electrode
and a complete interaction with the whole amount of \ce{Bi}.
These hypotheses are valid only in a 0D model. Nevertheless, they are used
in the following derivation due to the absence of a mass transport model
(describing the concentration distribution in the positive electrode). 
The electrochemical heat $\dot{Q}_\text{r}$ as a function of the molar flow rate of Li $\dot{n}_\text{Li}$ is 
\begin{linenomath*}
\begin{equation}
\dot{Q}_\text{r}  = \pm \underbrace{\frac{j \cdot A}{n_{\text{e}^\text{-}}\cdot
\text{F} }}_{\dot{n}_\text{Li}}(T \Delta\overline{S}_\text{Li} +
\frac{1-x_\text{Li}}{x_\text{Li}}  \Delta\overline{H}_\text{Bi}) \ ,
\end{equation}
\end{linenomath*}
in which $j$ denotes the absolute value of the current density, $A$ the cross sectional area of
the cell, $x_\text{Li}$ the molar fraction of lithium in bismuth,
$\Delta\overline{S}_\text{Li}$ the partial molar entropy of Li
and $\Delta\overline{H}_\text{Bi}$
the partial molar enthalpy of bismuth. The plus sign in the formulation refers to discharge and the minus refers to charge.
Positive values refer to heat that is absorbed, i.e. the cell cools down.\\
The electrochemical heat includes two terms.
The first term is the classical isothermal reversible heat
term $T\Delta\overline{S}_\text{Li}$, directly related to the
temperature coefficient of the cell equilibrium voltage
$\frac{\text{d}E_\text{cell,eq}}{\text{d}T}$ \cite{Bernardi1985}. The second one
is an additional term that takes into account the variation of enthalpy of
bismuth. The bismuth atoms are not directly affected by the
electrochemical reaction, but they mix with lithium atoms, so
they contribute to the heat generation term with their enthalpy
of mixing. The presence of this second term introduces non negligible
differences, as Tab.~\ref{tab:electrochemicalheat}
shows, but globally the electrochemical heat remains in
the same order of magnitude as the reversible one.   \\
If the electrochemical heat is divided by the
molar flow rate of Li $\dot{n}_\text{Li}$ (which is related to the electrical current through
the Faraday law) it is possible to define the electrochemical
heat $\overline{q}_\text{r,Li}$ per unit mole of Li:
\begin{linenomath*}
\begin{equation}
\label{eq:electrochemicalheat}
\overline{q}_\text{r,Li} = \frac{\dot{Q}_\text{r}}{\dot{n}_\text{Li}} = \pm  \Big(T
\Delta\overline{S}_\text{Li} + \frac{1-x_\text{Li}}{x_\text{Li}} 
\Delta\overline{H}_\text{Bi} \Big) =  \pm\frac{1}{x_\text{Li}} (T \Delta\overline{S} +
(1-x_\text{Li})  \Delta\overline{G}_\text{Bi}) \ ,
\end{equation}
\end{linenomath*}
in which $\Delta\overline{S}$ is the total entropy variation and 
$\Delta\overline{G}_\text{Bi}$ is the partial molar Gibbs free energy of bismuth.\\
The thermoneutral voltage $E_\text{TN}$ for LMBs is then defined as
\begin{linenomath*}
\begin{equation}
\label{eq:electrochemicalheat}
E_\text{TN}  =  - \frac{\Delta
\overline{G}_\text{Li}}{n_{\text{e}^\text{-}}\cdot \text{F}}  -
\frac{\overline{q}_\text{r,Li}}{n_{\text{e}^\text{-}}\cdot \text{F}}=
E_\text{cell,eq} - \frac{\overline{q}_\text{r,Li}}{n_{\text{e}^\text{-}}\cdot
\text{F} } \ .
\end{equation}
\end{linenomath*}
If the cell voltage $E_\text{cell}$ is equal to the thermoneutral voltage,
the heat generated by overpotentials (irreversibilities) is balanced by the
electrochemical heat. This is true only in point-like (0D) isothermal cells. \\
The values of the thermodynamic quantities ($\Delta \overline{G}_\text{Li}$,
$\Delta \overline{G}_\text{Bi}$, $\Delta \overline{H}_\text{Bi}$, $\Delta
\overline{S}$ and $\Delta \overline{S}_\text{Li}$) and the
equilibrium quantities ($E_\text{cell,eq}$, $\overline{q}_\text{r,Li}$,
$E_\text{TN}$) are calculated from the experimental
results of Gasior et al. \cite{Gasior1994}. They measured
the equilibrium cell voltage $E_\text{cell,eq}$  and its
temperature coefficient $\frac{\text{d}E_\text{cell,eq}}{\text{d}T}$
for a wide range of concentrations and temperatures close to the region
of interest. In order to avoid further
approximations due to extrapolation, the thermodynamic values
are evaluated at $T_\text{G}=\SI{775}{\kelvin}$ (the subscript G refers to
Gasior et al.). For the same reason, the fully
charged cell is studied with the values
at $x_\text{Li}=\SI{1}{\percent}$.
Numerical integration of the Gibbs-Duhem equation is used
in order to define all quantities of interest. All values
are collected in Tab.~\ref{tab:electrochemicalheat}. The equilibrium cell
voltage $E_\text{cell,eq}$ and the thermoneutral voltage $E_\text{TN}$ for the
fully charged state are shown  in Fig.~\ref{f:voltage} as black lines.
\begin{table}[h!]
\centering
\caption{Thermodynamic properties and electrochemical heat for three different charge states of a \ce{Li||Bi}
cell. Data calculated from Gasior et al. at $T_\text{G}=\SI{775}{\kelvin}$
\cite{Gasior1994}.}\label{tab:electrochemicalheat}
\begin{tabular}{llrrr}
\hline
\multicolumn{1}{c}{property}                                           &
\multicolumn{1}{c}{unit}                          & \multicolumn{3}{c}{cases studied} \\ \midrule
$x_\text{Li}$ & \si{\percent}                                     &    1 &   
10 &                   38 \\ \\[-1em]
$E_\text{cell,eq}$                                                     &
\si{\volt}                                        & 1.05 &  0.88 &              
0.73 \\ \\[-1em]
$\frac{\text{d}E_\text{cell,eq}}{\text{d}T}$                           &
\si{\micro\volt\per\kelvin}                          &  255 &    96 &           
$-$47.6 \\ \\[-1em]
$\Delta \overline{S}_\text{Li}$                                        &
\si{\joule.\mole}$^{-1}_\text{Li}$\si{\per\kelvin}     & 24.6 &   9.3 &         
$-$4.6 \\ \\[-1em]
$\Delta \overline{H}_\text{Bi}$                                        &
\si{\kilo\joule.\mole}$^{-1}_\text{Bi}$                &    0 & $-$0.19 &       
$-$1.0 \\ \\[-1em]
$T_\text{G}\Delta \overline{S}_\text{Li}$                              &
\si{\kilo\joule.\mole}$^{-1}_\text{Li}$                &   19 &   7.2 &         
$-$3.6 \\ \\[-1em]

$\Delta \overline{S}$                                                  &
\si{\joule.\mole}$^{-1}_\text{tot}$\si{\per\kelvin} & 0.33 &  1.83 &            
2.24 \\ \\[-1em]
$T_\text{G}\Delta \overline{S}$                                               
& \si{\kilo\joule.\mole}$^{-1}_\text{tot}$               & 0.25 &   1.4 &      
1.7 \\ \\[-1em]
$\Delta \overline{G}_\text{Bi}$                                        &
\si{\kilo\joule.\mole}$^{-1}_\text{Bi}$        &  $-$0.065	    & $-$0.78      &
$-$4.99 \\ \\[-1em]
$\overline{q}_\text{r,Li}$                                             &
\si{\kilo\joule.\mole}$^{-1}_\text{Li}$                &   19 &   5.5 &         
$-$5.2 \\  \\[-1em]
$E_\text{TN}$                                            & \si{\volt}         
& 0.86   &  0.82  &    0.79             \\ \\[-1em] \hline		

\end{tabular}\vspace{3pt}
\end{table}

\subsection{Polarization curve and parameters definition}\label{ch:polarisation}
The equilibrium cell potential $E_\text{cell,eq}$, 
and the study of classical overpotential 
theory allow to estimate the polarization curve of \ce{Li||Bi} LMBs. 
Knowing the polarization curve is crucial to define realistic
electrolyte layer thicknesses and current densities.

\begin{figure}[bht!]
\centering
{\includegraphics[width=1\textwidth]{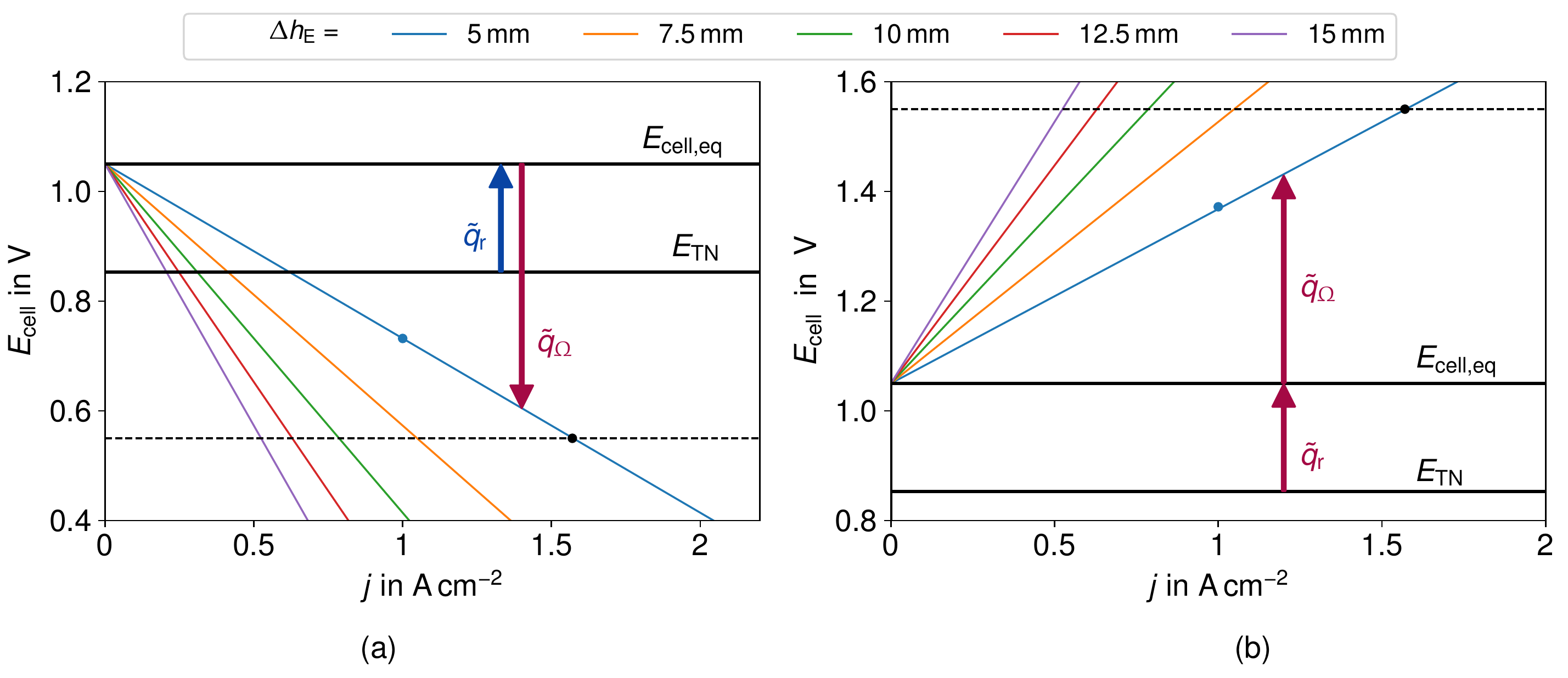}}\hfill
\caption{Polarization curve at $x_\text{Li}=\SI{1}{\percent}$
as function of the thickness of
the electrolyte ($\Delta h_\text{E}$), for discharge (a) and charge (b) according to Eq.~\ref{eqn:voltage}. The arrows are proportional to the heat
generated (red) or absorbed (blue) per unit of current. The ohmic term
($\tilde{q}_\Omega =-\rho_\text{el,E} \cdot j \cdot h_\text{E} $) is proportional
to current, while the electrochemical term ($\tilde{q}_\text{r} = \pm
\frac{\overline{q}_\text{r,Li}}{n_{\text{e}^\text{-}}\cdot \text{F} } $) is
constant. The blue dot marks the investigated condition with the fully 3D
simulation. The dashed line refers to the assumed maximum voltage drop of 
$\Delta E_\text{cell,$\Omega$} = \SI{0.5}{V}$. The black dot marks the 
maximum current density used in this study. }
\label{f:voltage}
\end{figure}

The fully-liquid interfaces between electrolyte 
and electrodes guarantee very fast charge transfer 
kinetics \cite{Kim2013b,Braunstein1971,Newhouse2017}, allowing 
to neglect the charge transfer overpotential. Furthermore, the 
fast diffusion in the liquid metal electrode ensures efficient mass 
transport during operation. For a first estimate it seems therefore reasonable to neglect the concentration 
overpotential \cite{Kim2013b,Braunstein1971}. 
The ohmic overpotential is consequently the most relevant 
inefficiency of LMBs. The simplified polarization 
curve $E_\text{cell}(j)$ is \cite{Shimotake1967}:
\begin{linenomath*}
\begin{equation}
\label{eqn:voltage}
E_\text{cell}(j) = E_\text{cell,eq}\pm \sum_{i}\eta_{i}(j)\cong 
E_\text{cell,eq} \pm \rho_\text{el,E} \cdot \Delta h_\text{E} \cdot j \ =
E_\text{cell,eq} \pm \Delta E_\text{cell,$\Omega$}\ ,
\end{equation}
\end{linenomath*}
in which $\rho_\text{el,E}$ is the electrical resistivity of the molten salt
electrolyte and  $\Delta h_\text{E}$ is the thickness of this layer; plus refers to the
charge of the cell, minus refers to the discharge.
The resulting linear profile is shown quite clearly in the report of Chum et
al. \cite{Chum1981} for the case of a \ce{Li||Bi} bimetallic cell.
\\
It is important to notice that the simplifications mentioned above
are not valid for all type of LMBs (e.g. Ca$||$Bi \cite{Kim2013a}) 
and operating conditions. For example, if the 
composition of the alloy locally (at the interface) reaches 
the one of the liquidus line, an additional nucleation 
overpotential occurs. In order to unveil all 
these complexities, the mass transport and the 
electrochemical phenomena must be introduced inside the model.\\
The presented polarization model does not depend on
the state of charge of the cell. It allows the direct
connection of two important parameters: the thickness of 
the electrolyte $\Delta h_\text{E}$ and the current density $j$ through the 
ohmic overpotential of the cell $\Delta E_\text{cell,$\Omega$}$. This 
connection will be used later in the heat conduction model.\\
The estimated polarization curves $E_\text{cell}(j)$, for charge and discharge, are shown in Fig.~\ref{f:voltage} as a function
of the thickness $\Delta h_E$ of the electrolyte layer.
The thermoneutral voltage divides the polarization plot (in discharge mode, Fig.~\ref{f:voltage}a) in
two regions. In the upper region ($E_\text{cell} > E_\text{TN}$), the negative
electrochemical heat (blue arrow) is predominant and the cell is
globally cooled (adiabatic conditions assumed). In the lower region of the plot
(in discharge mode,  Fig.~\ref{f:voltage}a) the heating due to the ohmic overpotential is the most
important and the cell is globally heated.
During charging (Fig.~\ref{f:voltage}b) both effects (electrochemical and ohmic) heat up the cell (red
arrows).
These considerations are valid only for the chosen $x_\text{Li}=\SI{1}{\percent}$, 
they strictly depend on the sign of the electrochemical 
heat per unit mole of \ce{Li} $\overline{q}_\text{r,Li}$.

\section{Cell geometry}\label{s:cellgeometry}
\begin{figure}[bht!]
\centering
{\includegraphics[width=1\textwidth]{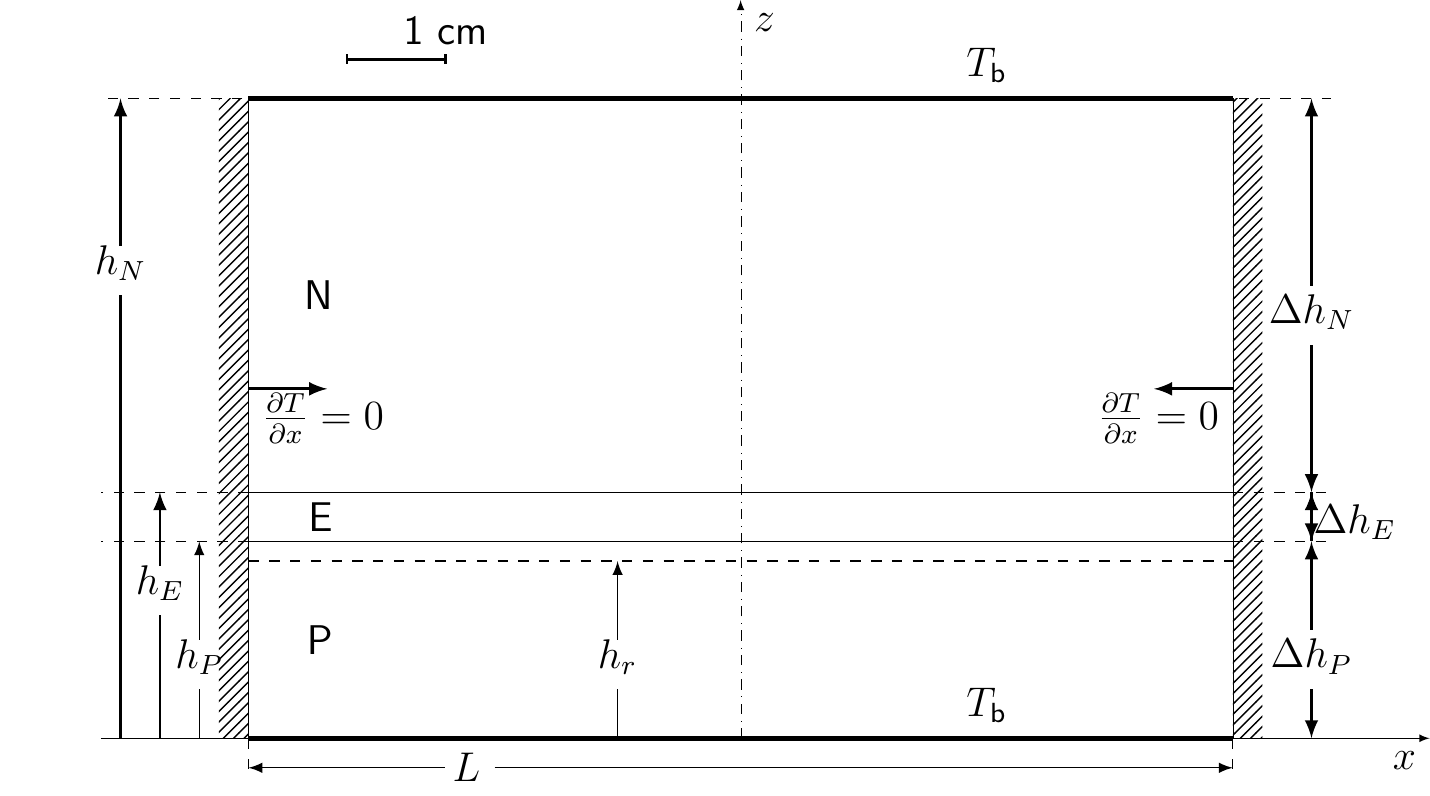}}\hfill
\caption{Cell dimensions and thermal boundary conditions.}
\label{f:geometry}
\end{figure}

We define a square-based cell with \SI{100}{\mm} side length $L$ and denote the
three phases from bottom to top as P (positive electrode), E (electrolyte) and N
(negative electrode), as shown in Fig.~\ref{f:geometry}.
In the first case, the fully charged cell is studied. The
height of the liquid bismuth layer is
$\Delta h_\text{P}=\SI{20}{\mm}$; the electrolyte thickness is
$\Delta h_\text{E}=\SI{5}{\mm}$. The maximum possible discharge capacity
of Li$||$Bi is given by the maximum molar fraction of lithium in the positive 
electrode, namely $x_\text{Li}=\SI{75}{\percent}$ \cite{Ning2015}. This results in a height of the lithium
layer of about $\Delta h_\text{N}=\SI{40}{\mm}$. 
For the partly discharged cells ($x_\text{Li}=\SI{10}{\percent}, \SI{38}{\percent}$),
the thickness of the liquid metal layers vary due to the amount 
of lithium transferred and the different densities of the alloys. These data 
are collected in Tab.~\ref{tab:charge_states}.
The values of the material properties of the components 
and their derivation are collected in \ref{ch:material_properties}.
\section{Thermal modeling}\label{s:thermalPhenomena}
\subsection{Thermal phenomena and main assumptions}\label{s:assumptions}
The temperature field in a liquid metal battery may change due to several
internal heat generation phenomena \cite{Bernardi1985, Newman2004}:
\begin{itemize} \itemsep0em
\item Joule heating induced by overpotentials
\item chemical and electrochemical reactions
\item phase changes
\end{itemize}
Furthermore, heat is transferred by conduction, advection and radiation 
as well as the Dufour effect (i.e. transport due to compositional gradients). Note that
mass transport and the chemical reactions induce composition variations
that may influence the thermodynamics (heat of mixing) and material
properties, but which are out of the current scope. Finally, the cell is a closed system that thermally
interacts with the environment and the thermal management system
\cite{Bradwell2016b}. In order to manage
properly this problem with the available data, several assumptions will
be made.

Ohmic heating of the electrolyte is usually the most important
overpotential in an LMB, as demonstrated in paragraph \ref{ch:polarisation}. 
However, the electrochemical heat released
when alloying Li into Bi may become important especially at low
charge or discharge currents \cite{Kim2013b}. We include (only) these
two heat sources in our models and neglect solidification. A uniform 
current distribution is assumed, ignoring MHD effects. The composition
of each layer is assumed to be constant and homogeneous.

Regarding the interaction with the external environment, we 
assume a simplified configuration similar to the one employed in 
previous studies \cite{Shen2015}. Only the three liquid layers 
are studied; the heat transfer in the gas layer and in 
the cell containment is neglected as well as radiative heat transfer (in the salt and argon layer).
The lateral walls are assumed adiabatic, whereas the bottom and the
top are considered isothermal at $T_\text{b} =\SI{450}{\celsius}$, 
as shown in Fig.~\ref{f:geometry}.
The cell is investigated in thermal steady state condition.
\subsection{One-dimensional heat conduction model}\label{sec:heat:conduction}
In this paragraph we study pure conduction only, because
it allows an analytical treatment. It provides a good approximation
for small fluid velocities and likewise an upper bound for the cell temperature.
For pure conduction and in a steady state limit, the temperature 
field depends only on the vertical coordinate $z$ as \cite{Carslaw1959}
\begin{linenomath*}
\begin{equation}
-k_i\frac{\text{d}^2 T}{\text{d} z^2}=\dot{q}'''_i 
\end{equation}
\end{linenomath*}
with $z$ denoting the vertical coordinate and $T$ the temperature; $k_i$
and $\dot{q}'''_i $ are the thermal conductivity and the heat
generation term of layer $i$. These three ordinary differential
equations are coupled with appropriate interface conditions: the
continuity of temperature and heat flux are enforced. We neglect
Joule heating in the metal layers because of their low electrical
resistivity.

\begin{figure}[h!]
\centering
\subfigure[]{\includegraphics[width=0.5\textwidth]{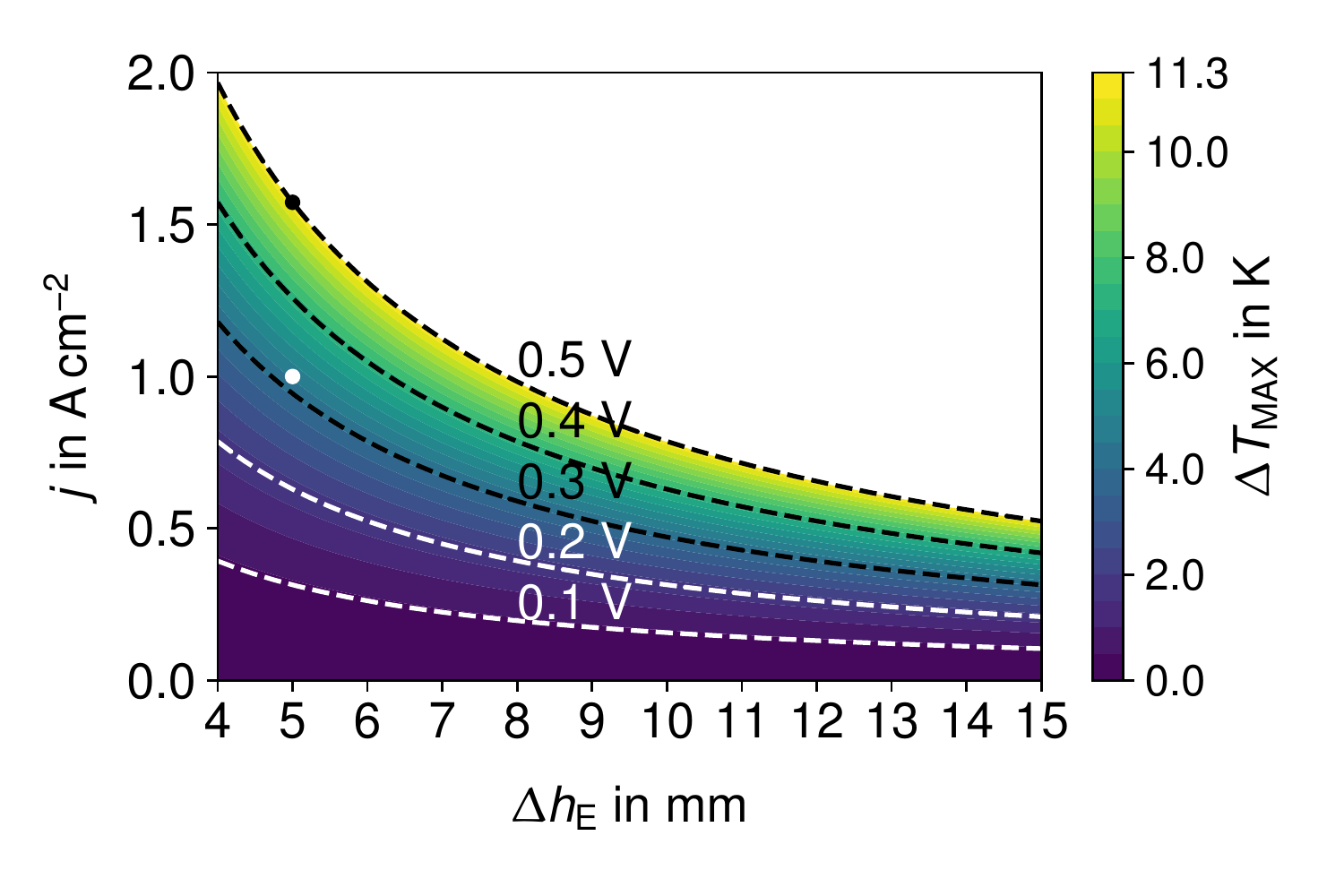}}\hfill
\subfigure[]{\includegraphics[width=0.5\textwidth]{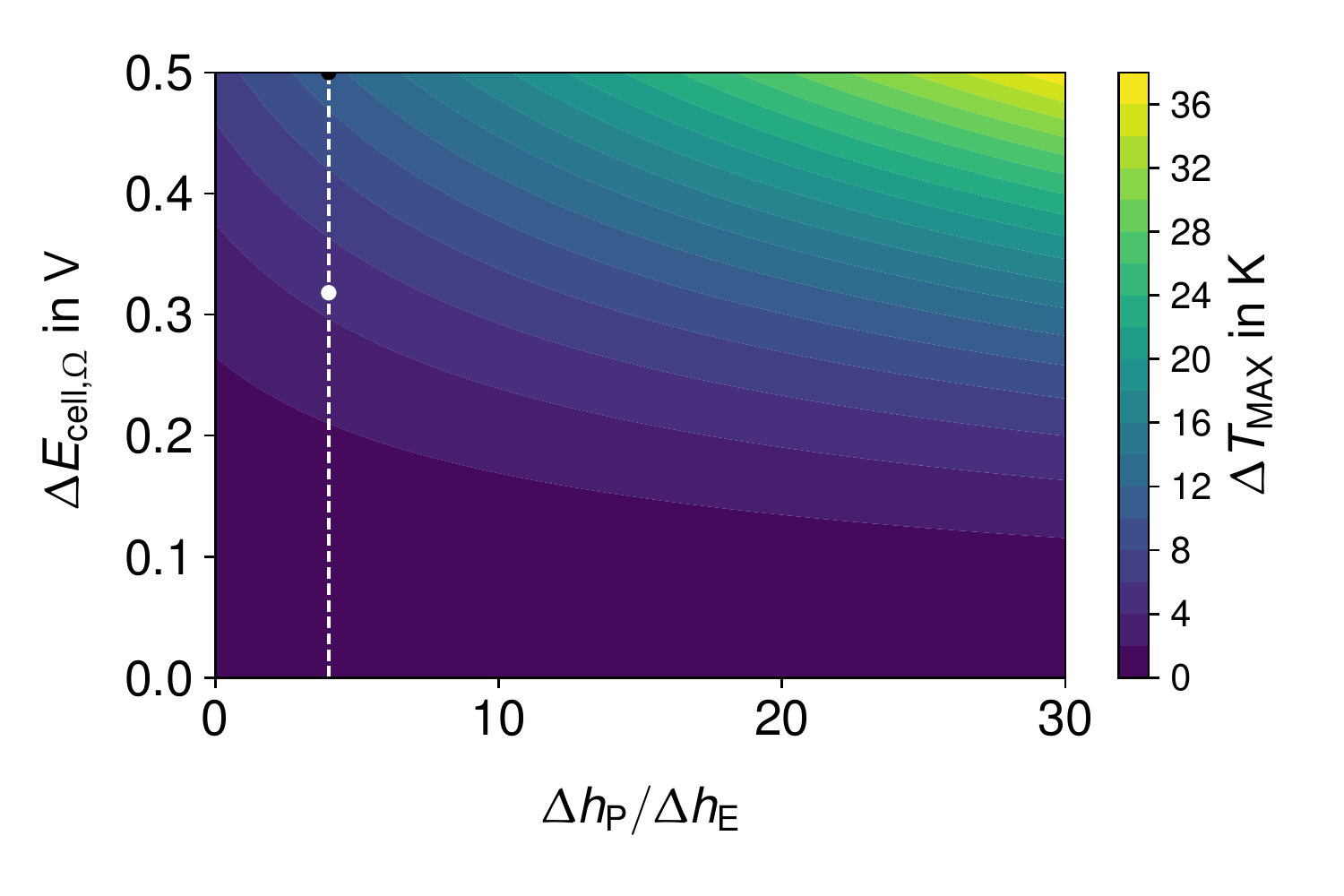}}\hfill
\subfigure[]{\includegraphics[width=0.5\textwidth]{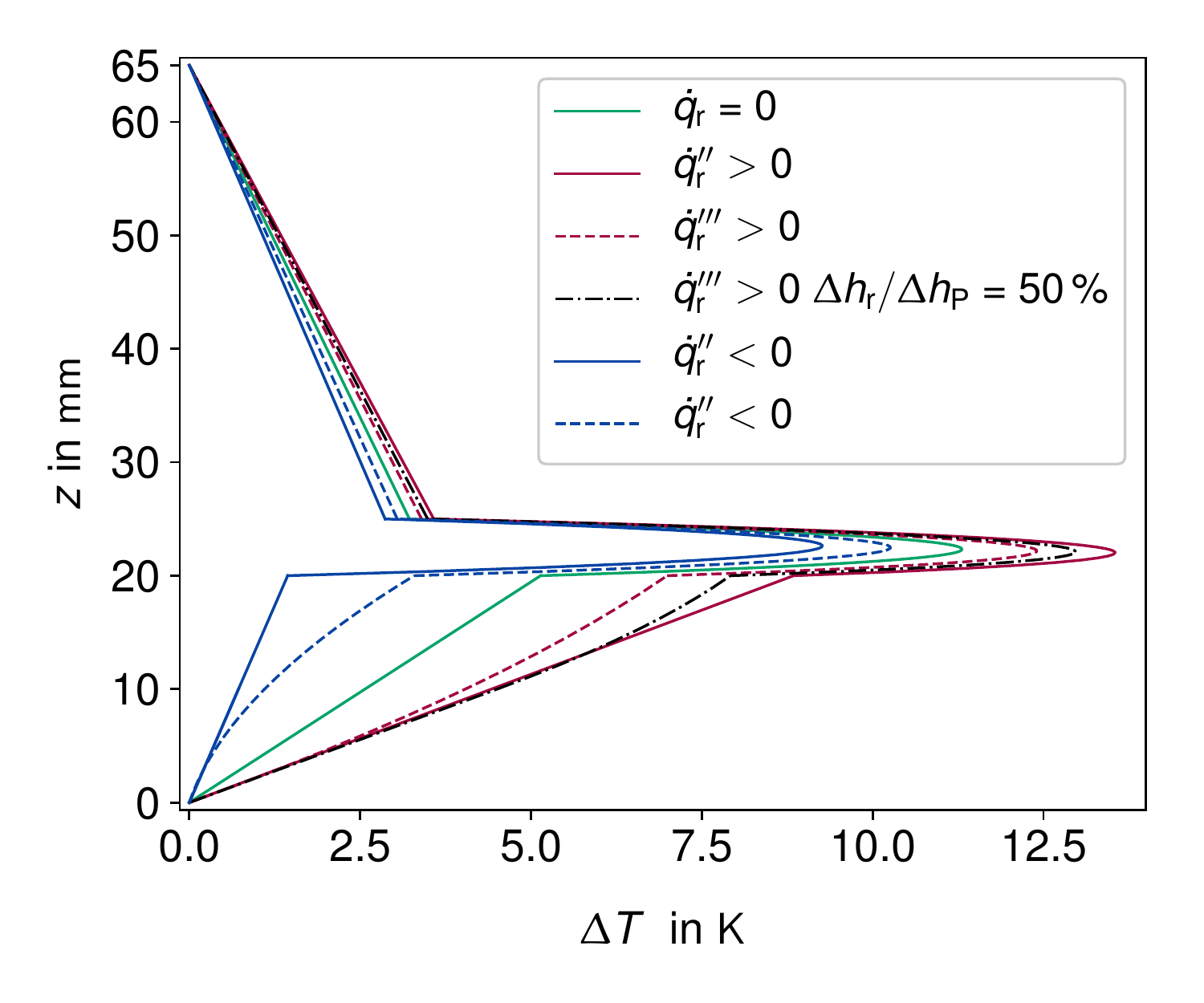}}\hfill
\subfigure[]{\includegraphics[width=0.5\textwidth]{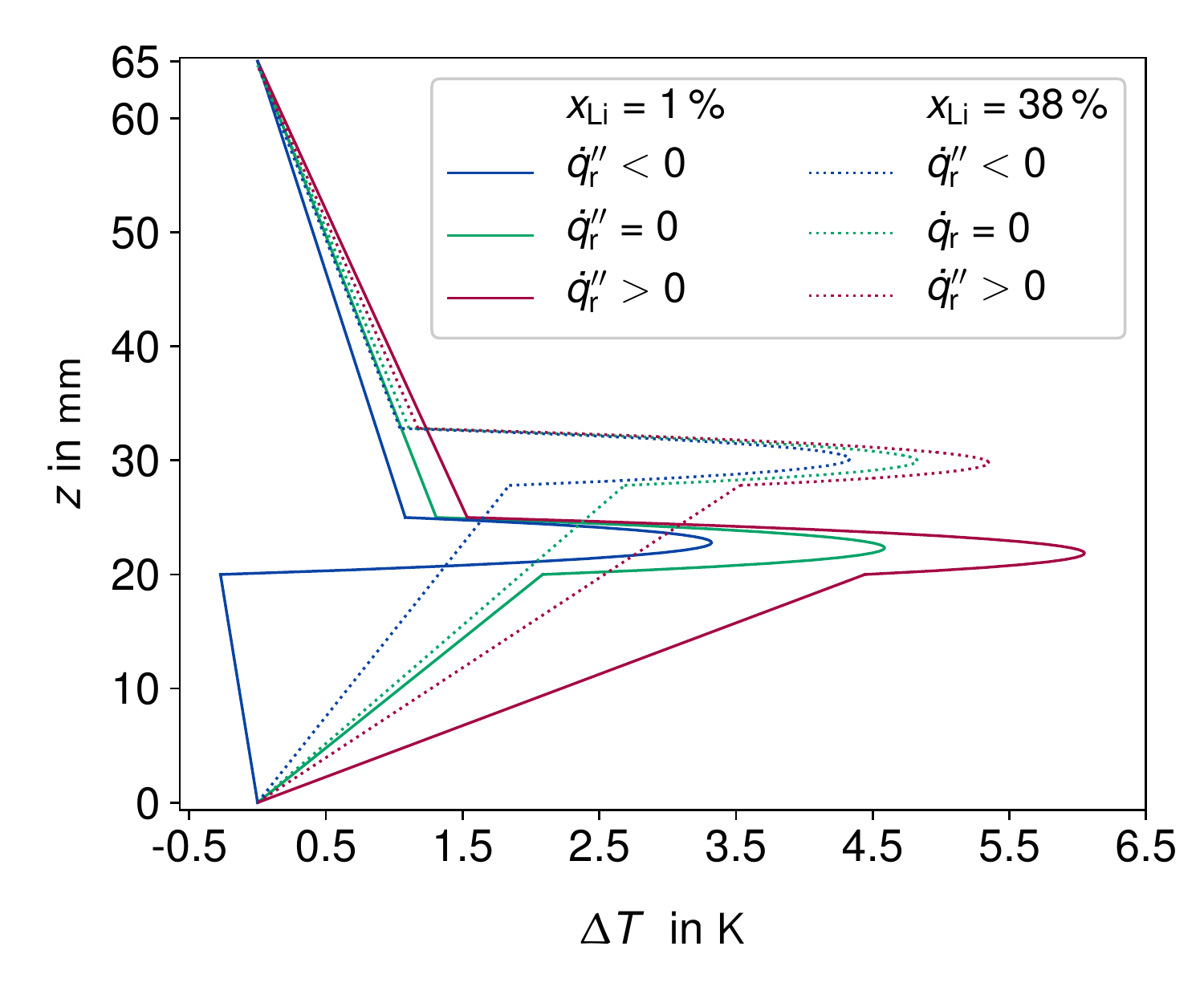}}\hfill
\caption{(a) Maximum cell temperature ($\Delta T_\text{max} = T_\text{max}- T_\text{b}$) as function of the electrolyte
thickness $h_\text{E}$ and the current density $j$ without
electrochemical heat generation $\dot{q}_\text{r}$. The white dot is the
investigated case with the fully 3D simulation at $x_\text{Li}=\SI{1}{\percent}$. The black dot
marks the maximum current density employed in the 1D study.
(b) Maximum cell temperature $\Delta T_\text{max}$ as function of $\Delta E_\text{cell,$\Omega$}$ and ratio
between the layer thickness of positive electrode and electrolyte $\Delta
h_\text{P}/\Delta h_\text{E}$, without
electrochemical heat generation $\dot{q}_\text{r}$. The white dashed line is for the geometry
employed at $x_\text{Li}=\SI{1}{\percent}$. 
(c) Comparison between vertical temperature distributions
in the presence and absence  of volumetric ($\dot{q}_\text{r}'''$ - dashed line) or
interfacial ($\dot{q}_\text{r}''$ - solid line) electrochemical heat generation. The
maximum current density is $j_\text{max} = \SI{1.57}{\ampere.\centi\metre^{-2}} $, 
the electrolyte thickness $h_\text{E} = \SI{5}{\milli\meter}$. (d)
Comparison between vertical temperature profiles in the presence and in absence
of interfacial ($\dot{q}_\text{r}''$)
electrochemical heat generation at the selected 
current density $j = \SI{1}{\ampere.\centi\metre^{-2}} $ 
for two different states of charge.}
\label{f:conduction}
\end{figure}

The electrochemical heat source is modeled
in two different ways: as a volumetric and interfacial effect. Both
approaches have been already employed for the study of the thermal behavior
of batteries \cite{Gu2000,Kumaresan2008,Newman2004,Min2012}, since
an exact treatment would require a complicated mass transfer model.
In the volumetric approach, the positive electrode is split into two parts.
The electrochemical heat is generated only in the active layer of
height $\Delta h_\text{r} = h_\text{P}-h_\text{r}$, and a homogeneous 
equation is solved in the inactive layer.
The volumetric heat generation reads:
\begin{linenomath*}
\begin{equation}
\label{eq:heatgeneration}
\dot{q}'''_i=
\begin{dcases}
0 & \text{for   $0 < z < h_\text{r}$\enspace\quad\text{inactive positive
electrode}}\\
\dot{q}_\text{r}'''= -\frac{j
\overline{q}_\text{r,Li}}{n_{\text{e}^\text{-}}\text{F} \Delta h_\text{r}}& \text{for
$h_\text{r} < z < h_\text{P}$\quad\text{active positive electrode}}\\
\dot{q}_{\Omega, \text{E}}''' = \rho_\text{el,E} j^2 & \text{for $h_\text{P} < z <
h_\text{E}$\quad\text{electrolyte}}\\
0& \text{for $h_\text{E} < z < h_\text{N}$\quad\text{negative electrode}} \ .
\end{dcases}
\end{equation}
\end{linenomath*}
In the interfacial model, the electrochemical heat generation
is imposed on the interface between the positive electrode
and the electrolyte ($z = h_\text{P} $) as a discontinuity of the heat flux
(this approximation is also employed for Peltier heat 
\cite{Deen1998}): 
\begin{linenomath*}
\begin{equation}
-k_\text{E}\frac{\text{d}T}{\text{d}z}\Bigr|_{z =
h_\text{P}+\epsilon}+k_\text{P}\frac{\text{d}T}{\text{d}z}\Bigr|_{z =
h_\text{P}-\epsilon}=\underbrace{-\frac{j
\overline{q}_\text{r,Li}}{n_{\text{e}^\text{-}}\text{F}
}}_\text{$\dot{q}_\text{r}''$} \ ,
\end{equation}
\end{linenomath*}
with $\epsilon$ denoting an infinitesimal distance.
The solution of the general formulation that takes into account both
electrochemical heat generation models is expressed by the following temperature
distribution:
\begin{linenomath*}
\begin{eqnarray}
T=
\begin{dcases}
c_1\cdot z + c_2 & \text{for   $0 < z < h_0$\enspace\quad\text{inactive positive
electrode}}\\
-\frac{\dot{q}_\text{r}'''}{2k_\text{P}}\cdot z^2 + c_3 z + c_4 & \text{for $h_0
< z < h_\text{P}$\quad\text{active positive electrode}}\\
-\frac{\dot{q}_{\Omega, \text{E}}'''}{2k_\text{E}}\cdot z^2 + c_5 z + c_6 & \text{for
$h_\text{P} < z < h_\text{E}$\quad\text{electrolyte}}\\
c_7\cdot z + c_8 & \text{for $h_\text{E} < z < h_\text{N}$\quad\text{negative
electrode}} \\ 
\end{dcases}
\end{eqnarray}
\end{linenomath*}
with the parameters $c_i$ given in \ref{ch:constants_conduction}, where it
is understood that either $\dot{q}_\text{r}'''$ or $\dot{q}_\text{r}''$ vanishes.
The maximum cell temperature $T_\text{max}$ and its position $z_\text{max}$ are
\begin{linenomath*}
\begin{equation}
T_\text{max} = \frac{k_\text{E} c_5^2}{2\dot{q}_\Omega'''} + c_6 \  \
\text{at} \ \ z_\text{max} = \frac{ k_\text{E}c_5}{\dot{q}_\Omega'''} \ ,
\end{equation}
\end{linenomath*}
with the maximum value being a parabolic function of the current density.
In Fig.~\ref{f:conduction} we present some results derived from the
analytical model. Fig.~\ref{f:conduction}a shows the maximum temperature
difference in the cell for typical current densities and electrolyte thicknesses
during discharge, and without electrochemical heat generation. Assuming a fixed
ohmic drop $\Delta E_\text{cell,$\Omega$}$, it is possible to relate the current
density to the thickness of the electrolyte by applying Eq.~\ref{eqn:voltage}.
This allows to bound the current and electrolyte thickness to an admissible region.
If the ratio between the thickness of the layers is fixed (Fig.~\ref{f:conduction}a
with $\Delta h_\text{P}/\Delta h_\text{E} = 4$ and $\Delta h_\text{N}/\Delta h_\text{P} =2$), the curves for 
constant $\Delta T_\text{max}$ and constant $\Delta E_{\text{cell},\Omega}$ coincide.

Starting from these consideration it is possible to derive the results presented
in Fig.~\ref{f:conduction}b. It shows the maximum temperature difference as a
function of the ohmic overpotential  and the ratio
$\Delta h_\text{P}/\Delta h_\text{E}$, at fixed $\Delta h_\text{N}/\Delta
h_\text{P} = 2$. The temperature rises up in the cell due to the increase of the ohmic voltage drop and due to the increase of the thickness
of the liquid metal layers.

Fig.~\ref{f:conduction}c shows the vertical temperature profile in the fully charged cell
at the maximum current density ($j_\text{max} =
\SI{1.57}{\ampere.\centi\metre^{-2}} $). Positive (red), negative (blue) and
null (green) electrochemical heat generation are presented. The temperature profiles in
the top electrode and in the electrolyte are always linear and parabolic,
respectively.  
The temperature profile in the lower electrode
is linear in case of interfacial (blue and red solid lines) or null (green) electrochemical heat generation. However, in presence of the volumetric effect the profile becomes
parabolic. As an example, the black dotted line shows the case where the electrochemical heat generation
is released in the upper half part of the positive electrode 
($\Delta h_\text{r}/\Delta h_\text{P} = \SI{50}{\percent}$). As we see from Fig.~\ref{f:conduction}c, 
we obtain always the strongest effect if we assume the 
electrochemical heat generation to be released at the interface (and not in the volume).

The electrochemical heat generation becomes more and more important when we reduce the
current density, as already expected from Fig.~\ref{f:voltage}a and shown in Fig.~\ref{f:conduction}d.
For the fully charged cell at \SI{1}{\ampere.\centi\meter^{-2}} and in presence of
electrochemical cooling, the temperature profile in the
lower electrode has a negative slope. This may induce an unstable density
stratification -- and lead to convective flow if the Rayleigh number exceeds the critical value. More details, including the voltage at which the lower layer becomes isothermal, are presented in \cite{Personnettaz2017}.
Finally, Fig.~\ref{f:conduction}d shows also that the charging (or discharging of the
battery) leads to a shift of the thermal peak due to the change of thickness
of the liquid metal layers.
\subsection{Thermo-fluid dynamic multiphase solver and discretization}
In an LMB the transport of heat is not only governed by conduction
(so far considered only), but also to a considerable extend by advection.
In the following model, fluid flow is generated by thermally driven buoyancy forces,
while we neglect Marangoni and MHD effects due to their supposedly smaller
magnitude. Besides this source of momentum, momentum is transferred between the layers due to the
mechanical coupling and by the interaction with the containment
walls. The temperature dependence of the density of the fluids
is described by a linear law ($\rho_i =
\rho_\text{ref,$i$}(1-\beta_i(T-T_\text{ref}))$), in which
$\rho_\text{ref,$i$}$ is the density of the $i$-th fluid calculated at
the reference temperature $T_\text{ref}$ of \SI{450}{\celsius} and $\beta_i$ is
the volumetric thermal
expansion coefficient.  The
density variations due to temperature changes are sufficiently small
to assume an incompressible flow. The other properties are assumed to be
constant with respect to the temperature.

At the walls we apply homogeneous Dirichlet boundary condition for
both velocity components (no-slip and impermeable conditions). While
previous studies \cite{Shen2015, Koellner2017} used a multi-region 
approach, we employ a continuum field formulation with a homogeneous 
flow model \cite{Worner2003} to study the fluid-dynamics of the cells.
This is done in accordance to Weber et al. \cite{Weber2017},
in order to include MHD capabilities in a future development.
The model is implemented in the finite volume (FVM) library OpenFOAM
4.0 \cite{Weller1998} as an extension of the standard solver
\emph{multiphaseInterFoam}.

The incompressible Navier-Stokes equation (NSE) for a Newtonian fluid
is reformulated in order to take into account the multiphase nature of
the system \cite{Rusche2002} as
\begin{linenomath*}
\begin{equation}\label{eqn:nse}
\frac{\partial(\rho \bi u)}{\partial t} + \nabla\cdot(\rho\bi u\bi u)
= -\nabla p_\text{d} +  \text{g} z \nabla\rho +
\nabla\cdot(\rho\nu(\nabla\bi u + (\nabla\bi u)^\intercal)) 
+ \bi f_\text{st}
\end{equation}
\end{linenomath*}
with $\rho$ denoting density, $\bi u$ velocity, $t$ time,
$\text{g}$ gravity acceleration, $z$ the axial coordinate, $\nu$ kinematic viscosity
and $\bi f_\text{st}$ the source of momentum due to the surface tension. 
For the derivation of the modified pressure $p_\text{d}$, see \cite{Rusche2002,
Weber2017a}. \\
The interface capturing is done with the  volume of fluid method (VOF), which
uses a (volumetric) phase fraction $\alpha_i$ for each fluid $i$.
The advection of these quantities \cite{Ubbink1997,Rusche2002},
\begin{linenomath*}
\begin{equation}
\frac{\partial\alpha_i}{\partial t} + \nabla\cdot(\alpha_i\bi u ) = 0 \ ,
\end{equation}
\end{linenomath*}
together with the continuity equation $\nabla\cdot \bi u = 0$ guarantees the
conservation of mass. \\
The surface tension is implemented (using the CSF model of Brackbill et al.
\cite{Brackbill1992,Ubbink1997,Rusche2002,Kissling2010}) as a volumetric force
near the
interface as $\bi f_\text{st} = \sum_i\sum_{j\neq
i}\gamma_{ij}\kappa_{ij}\delta_{ij}$ with $\gamma_{ij}$ denoting the
(constant) interface tension between the phases $i$ and $j$. Concentration and
temperature Marangoni effects are completely neglected. The
curvature of the interface $i|j$ is
\begin{linenomath*}
\begin{equation}\label{eqn:kappa}
\kappa_{ij}=-\nabla\cdot\frac{\alpha_j\nabla\alpha_i -
\alpha_i\nabla\alpha_j}{|\alpha_j\nabla\alpha_i -
\alpha_i\nabla\alpha_j|}.
\end{equation}
\end{linenomath*}
The term $\delta_{ij}=\alpha_j\nabla\alpha_i -
\alpha_i\nabla\alpha_j$  applies the volumetric force only near the
interfaces, where variations of the indicator function are present. Thus, the
surface tension force becomes 
\begin{linenomath*}
\begin{equation}\label{eqn:fst}
\bi f_\text{st} = -\sum_i\sum_{j\neq i}\gamma_{ij}
\nabla\cdot\left(\frac{\alpha_j\nabla\alpha_i -
\alpha_i\nabla\alpha_j}{|\alpha_j\nabla\alpha_i -
\alpha_i\nabla\alpha_j|}\right)(\alpha_j\nabla\alpha_i -
\alpha_i\nabla\alpha_j).
\end{equation}
\end{linenomath*}
According with the hypotheses stated in section \ref{s:thermalPhenomena} 
the energy equation becomes:
\begin{linenomath*}
\begin{equation}
c_p \left(\frac{\partial\rho T}{\partial t} + \nabla\cdot(\rho T \bi u)\right) =
\nabla\cdot k\nabla T +\dot{q}_{\Omega}''' + \dot{q}_\text{r}''',
\end{equation}
\end{linenomath*}
with $c_p$ denoting the specific heat capacity. The two last terms
represent the ohmic and the electrochemical heat generation, which
were already specified in Eq.~\ref{eq:heatgeneration}. Now we consider
also the ohmic heat generation in the liquid metal layers, even if its
effect is negligible. The electrochemical heat generation is
implemented as a volumetric term $\dot{q}_\text{r}'''$ which
affects only a region of the lower layer of height $\Delta h_\text{r}$, just
below the interface.

The thermodynamics and transport properties are computed in each
computational cell using the volumetric phase fraction as a weighting
factor as 
\begin{linenomath*}
\begin{gather}
\begin{split}
\nu  =\frac{1}{\rho} \sum_i \alpha_i\rho_i\nu_i \quad
c_p = \frac{1}{\rho} \sum_i \alpha_i\rho_i c_{p,i}\quad
k = \Big(\sum_i \frac{\alpha_i}{k_i}\Big)^{-1}\\[1em]
\rho_\text{el} = \sum_i \alpha_i\rho_{\text{el,}i} \quad \text{and}
\quad
\rho = \sum_i \alpha_i\rho_\text{ref,$i$}(1-\beta_i(T-T_\text{ref})).
\end{split}
\end{gather}
\end{linenomath*}
The linear average is used for density and harmonic interpolation for
the conductivities. Specific heat capacity and kinematic viscosity are linearly weighted
by mass. For a detailed discussion of different blending schemes and their
applicability, please refer to \cite{Ubbink1997,Carson2005,Kumar2014,Nabil2016}.
Second order accurate schemes are used for time (backward scheme) and
space discretization (linear scheme). The advection of temperature is
discretized using the linearUpwind scheme while the bounded LUST scheme 
is used for velocity and vanLeer for the phase fractions 
\cite{Greenshields2016}.
The time step is determined using a Courant number of 0.5 
based on the fluid velocity, interface displacement and
capillary velocity; for more details see \ref{ch:spurious}. 

\subsection{Comparison with pseudo-spectral code}
The OpenFOAM solver is validated by comparison with an established code
\citep{Koellner2017} which uses a pseudo-spectral discretization. This reference
solver was already successfully applied to a variety of solutal and thermal convection
problems in two \citep{Boeck2002,Boeck2003,Koellner2014,Koellner2016} and
recently three-layer systems \citep{Koellner2017}. Its spatial discretization
relies on the expansion of fields into Fourier modes for the x-y directions and
Chebychev polynomials in the z direction. This type of discretization makes the
code limited by means of geometry, namely only periodic boundary conditions on
the side walls and non-deformable interfaces can be handled; but on the other
hand, it features a high accuracy and small numerical costs. The recent
developments of the code to capture three layers \citep{Koellner2017} have been
validated by reproducing linear stability results, reproducing laminar flow,
pure thermal conduction, as well as  checking the kinetic energy balance. 
\begin{figure}[h!]
	\centering
	\subfigure[]{\includegraphics[width=0.45\textwidth]{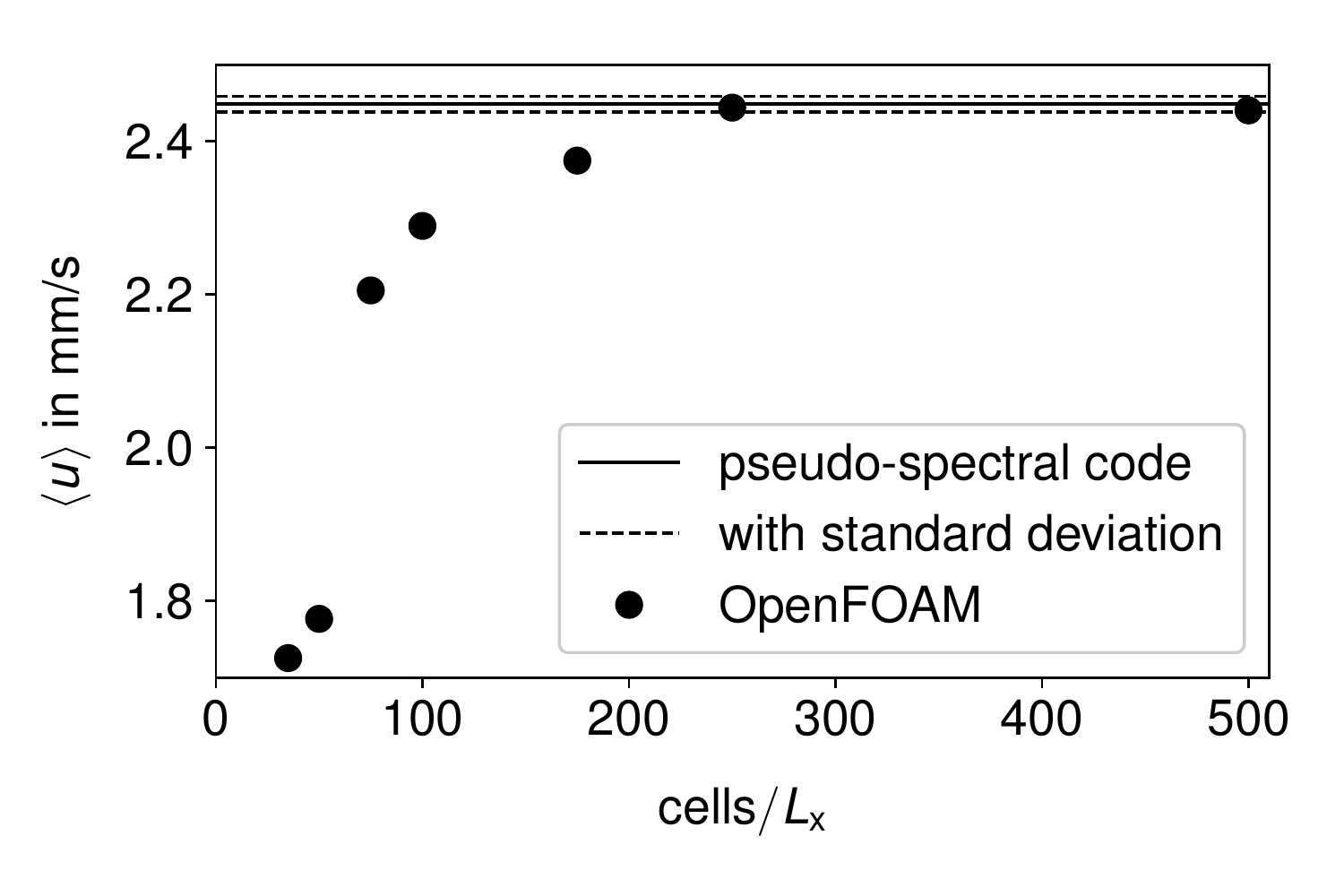}}\hfill
	\subfigure[]{\includegraphics[width=0.45\textwidth]{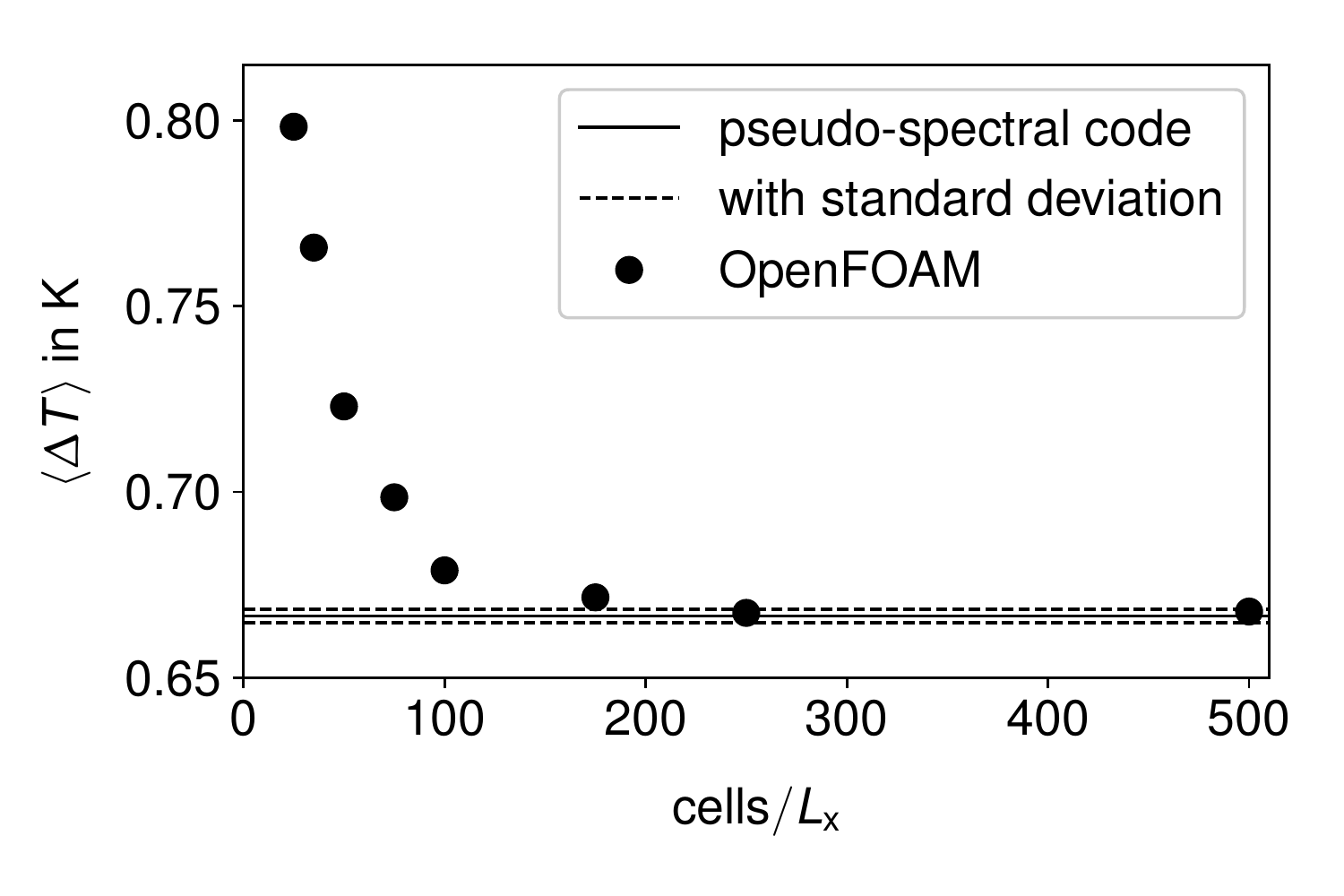}}\hfill
	\subfigure[]{\includegraphics[width=0.45\textwidth]{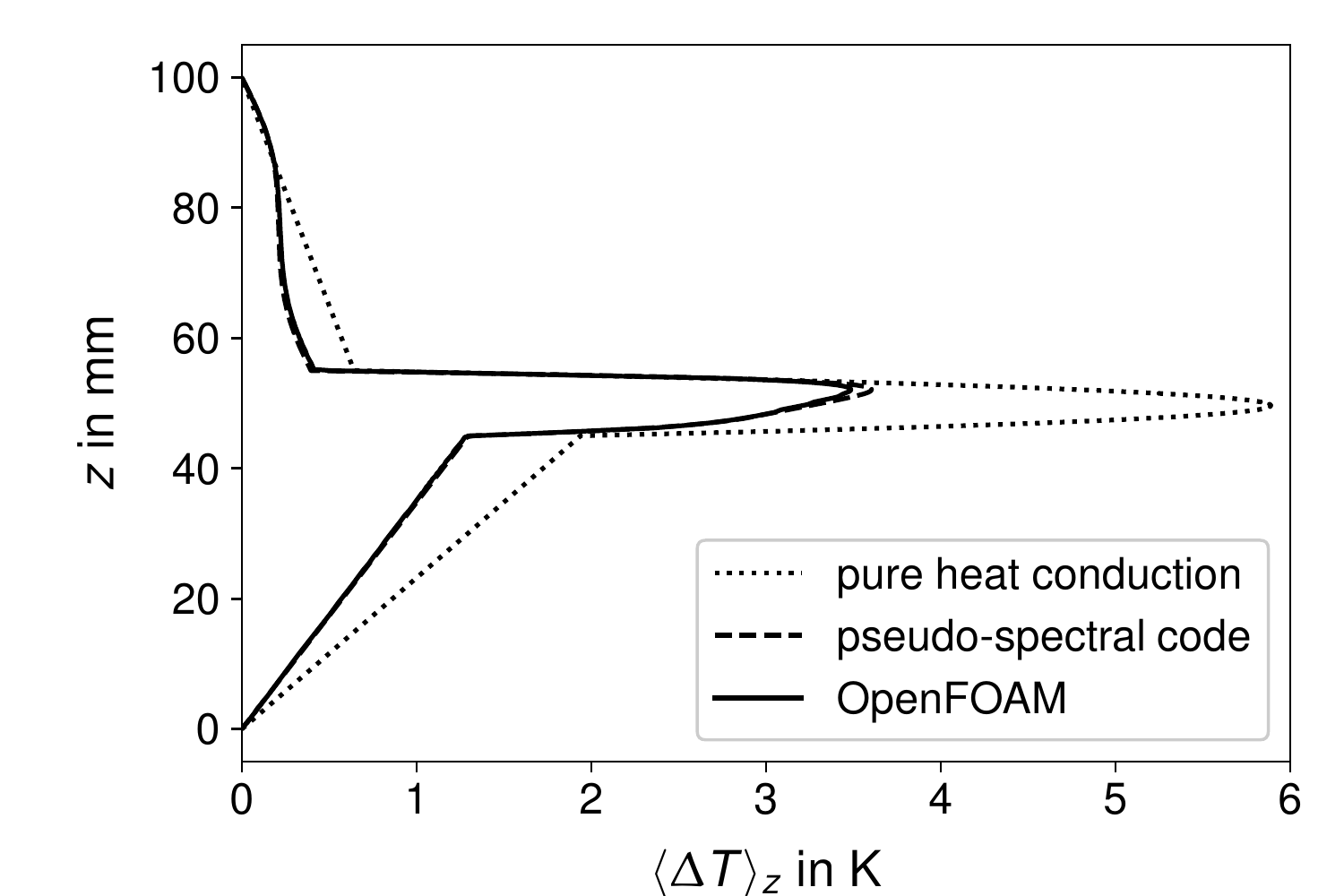}}\hfill
	\subfigure[]{\includegraphics[width=0.45\textwidth]{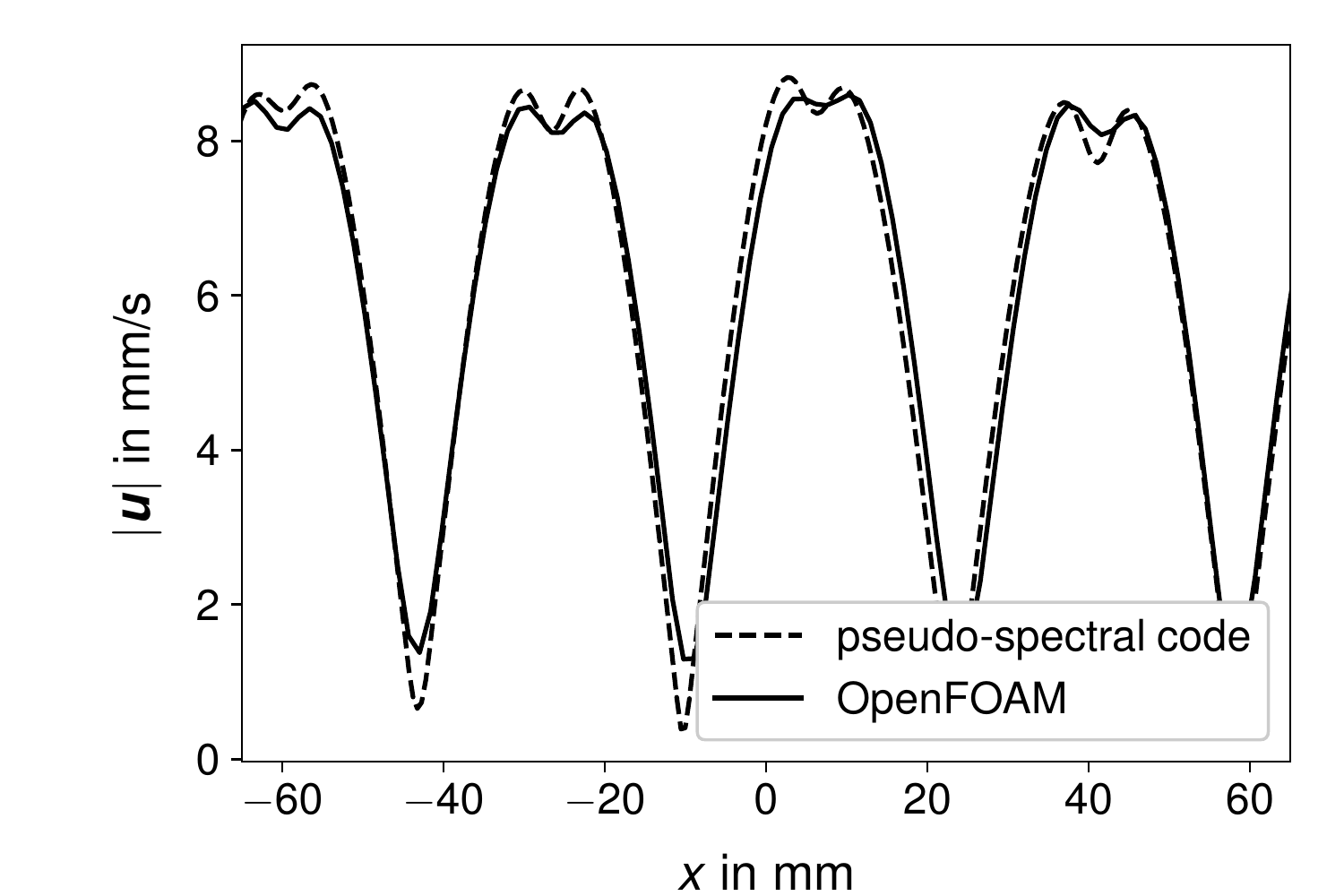}}\hfill
	\subfigure[]{\includegraphics[width=0.45\textwidth]{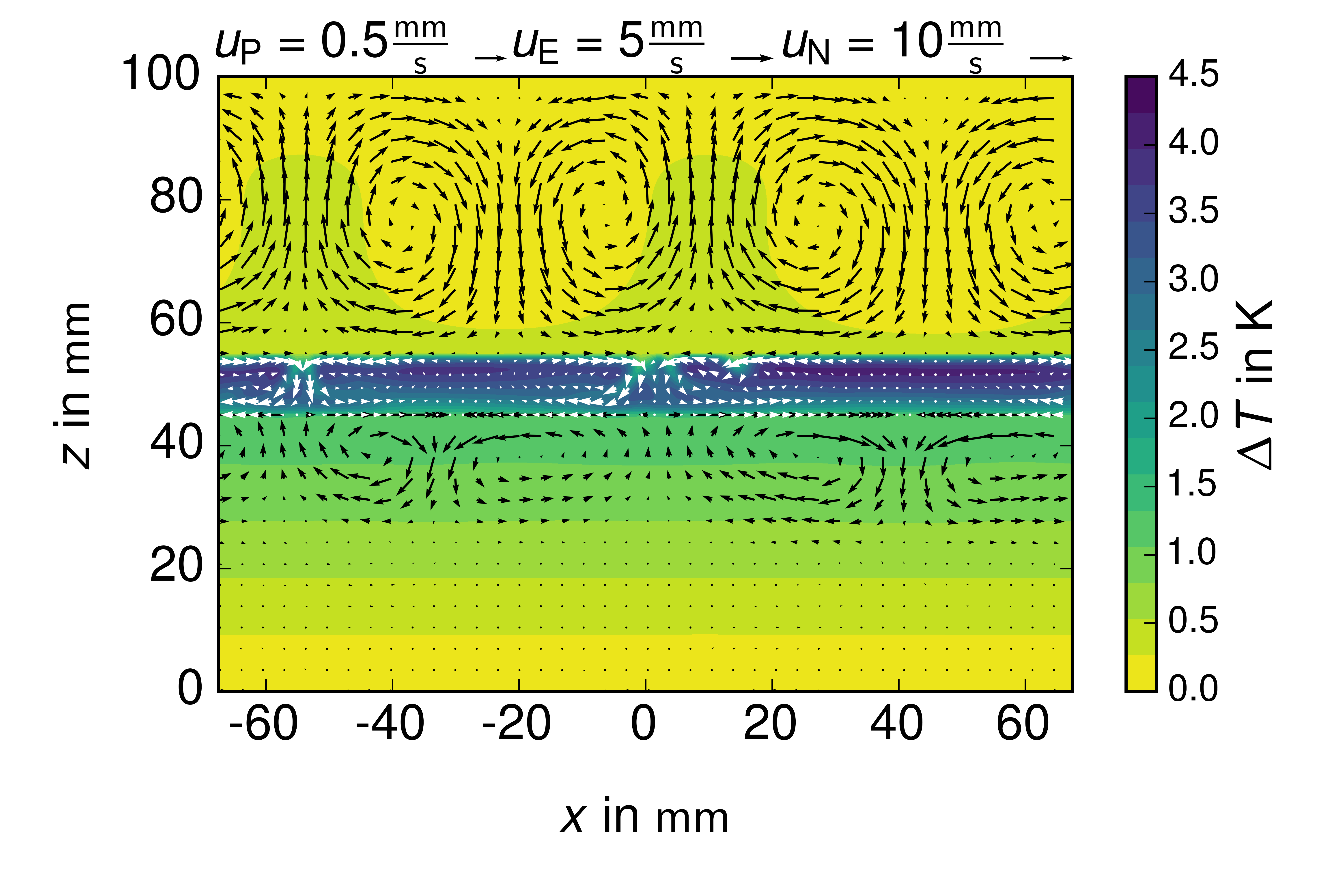}}\hfill
	\subfigure[]{\includegraphics[width=0.45\textwidth]{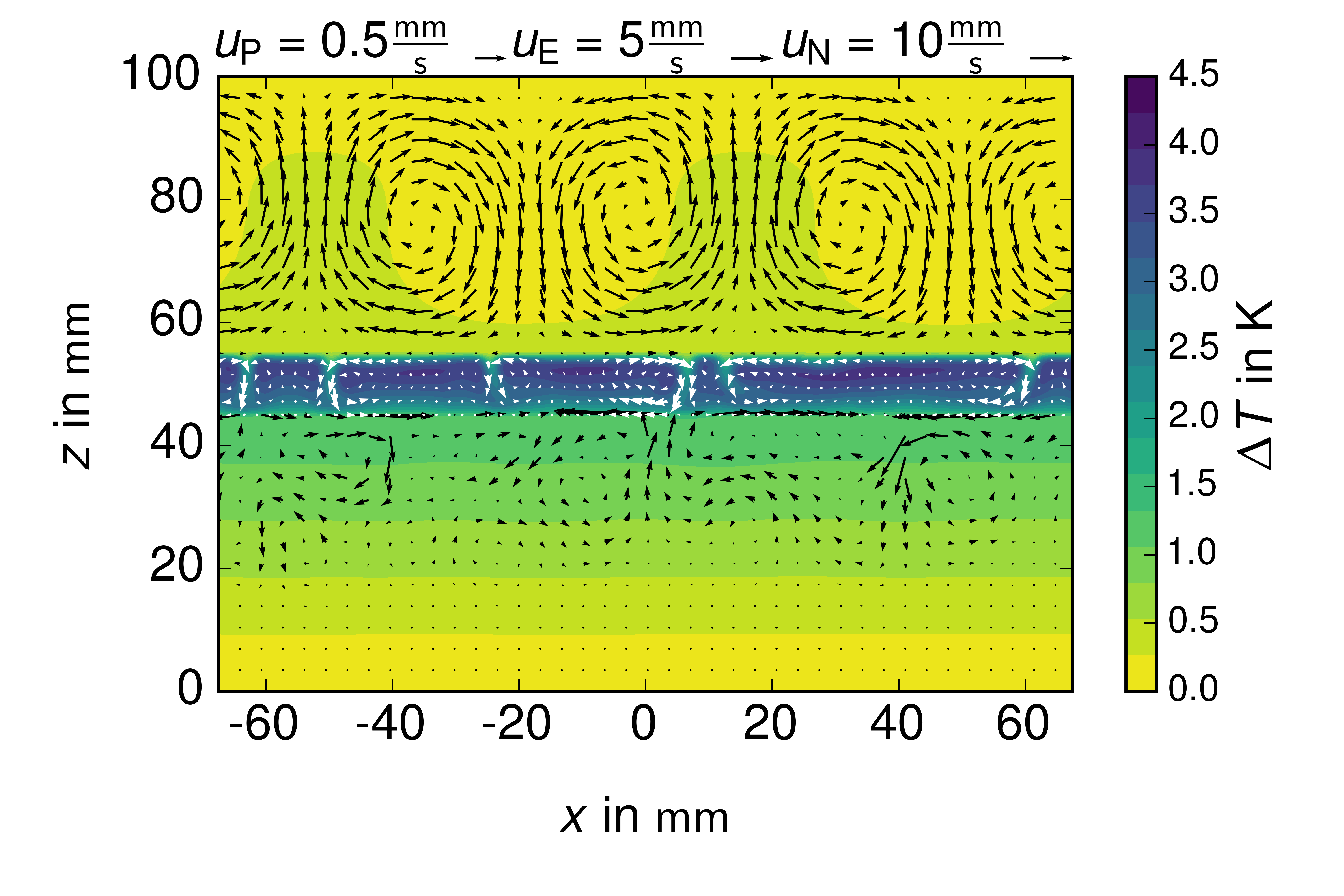}}\hfill
	\caption{Grid study for volume averaged velocity (a) and volume averaged
		temperature (b), mean temperature profile along the vertical axis (c) and local
		velocity profile along a horizontal line (d) (at $z = \SI{72.5}{\milli\meter}$) for the OpenFOAM solver and
		the spectral code. The general flow structure and temperature distribution of
		the pseudo-spectral (e) and the OpenFOAM solver (f). Note that we
		adapted the size of the velocity vectors for each phase with the
		reference values given in a legend above the subfigures.}
	\label{f:validation}
\end{figure}
The following 2D test case is used for comparison:
The boundary condition at the side walls are assumed to be periodic,
the temperature of top and bottom is set to  \SI{500}{\celsius}.
The material properties of K\"ollner et al. \cite{Koellner2017} are employed.
Furthermore, different dimensions ($L_\text{x}=\SI{135}{\milli\meter}$, 
$\Delta h_\text{P} =\Delta h_\text{N}=\SI{45}{\milli\meter}$,
$\Delta h_\text{E}=\SI{10}{\milli\meter}$) as shown in
Fig.~\ref{f:validation} are used, and the current density is set to
$j = \SI{0.5}{\ampere.\centi\meter^{-2}} $.
A grid study yields converged results of velocity and temperature when
using at least 250 cells ($N_\text{x}=250$) in the horizontal and
200 cells in the vertical direction ($N_\text{z}=200$). The mesh is strongly
refined in the region of the electrolyte, near the interfaces and the wall. 

For $N_x>250$, the global mean velocity deviates less than \SI{2}{\percent} and 
temperature by less than \SI{1}{\percent} between the two solvers.
The comparison is performed in thermal
steady state condition; the profiles are time averaged over a period of
\SI{50}{\second}, because the flow field is already in time-dependent, chaotic state.
The horizontally averaged temperature in Fig.~\ref{f:validation}c and the local
velocity magnitude in Fig.~\ref{f:validation}d (located in the middle of the
light metal at $z=\SI{72.5}{\milli\meter}$)
agree very well with the reference pseudo-spectral solution. A snapshot
of the velocity and temperature field is displayed for the pseudo-spectral
code in Fig.~\ref{f:validation}e and for the OpenFOAM solver in (f) for qualitative
comparison. The OpenFOAM solver captures especially well
the flow structure and the temperature distribution in the top layer. Although the 
velocity magnitude in the two other layers is comparable between the two solvers, the flow structure shows
some differences. This might have several reasons, as e.g. a less stable 
velocity distribution in the molten salt or the presence of spurious 
currents that affect the weak flow field in the lower layer. 

Generally, a very good agreement was found between the OpenFOAM and pseudo-spectral code. 
This is a good result, because both approaches are very different. 
For further details about the grid study and the
comparison, please refer to \cite{Personnettaz2017}. 
\subsection{Results}\label{s:results}
The OpenFOAM solver is used to study the thermo-fluid dynamics of the cell in
three states of charge (see section \ref{s:cellgeometry}). We
employ a current density of \SI{1}{\ampere.\centi\meter^{-2}} and study both,
the charge and discharge of the cell, with and without
electrochemical heat generation. The latter is prescribed as a volumetric effect in the upper
\SI{10}{\percent} of the thickness of the positive electrode ($\Delta h_r = 
\frac{ \Delta h_\text{P}}{10}$).\\
The LMB domain is discretized with an unstructured orthogonal
mesh. The latter is strongly
refined in the electrolyte and near the walls in order to resolve local
gradients, and near the interfaces in order to minimize numerical 
smearing. We employ 100 cells in horizontal and 75 cells in vertical
direction with a total number of $\SI{7.5e5}{}$ cells.\\
\begin{figure}[h!]
	\centering
	\subfigure[]{\includegraphics[width=0.48\textwidth]{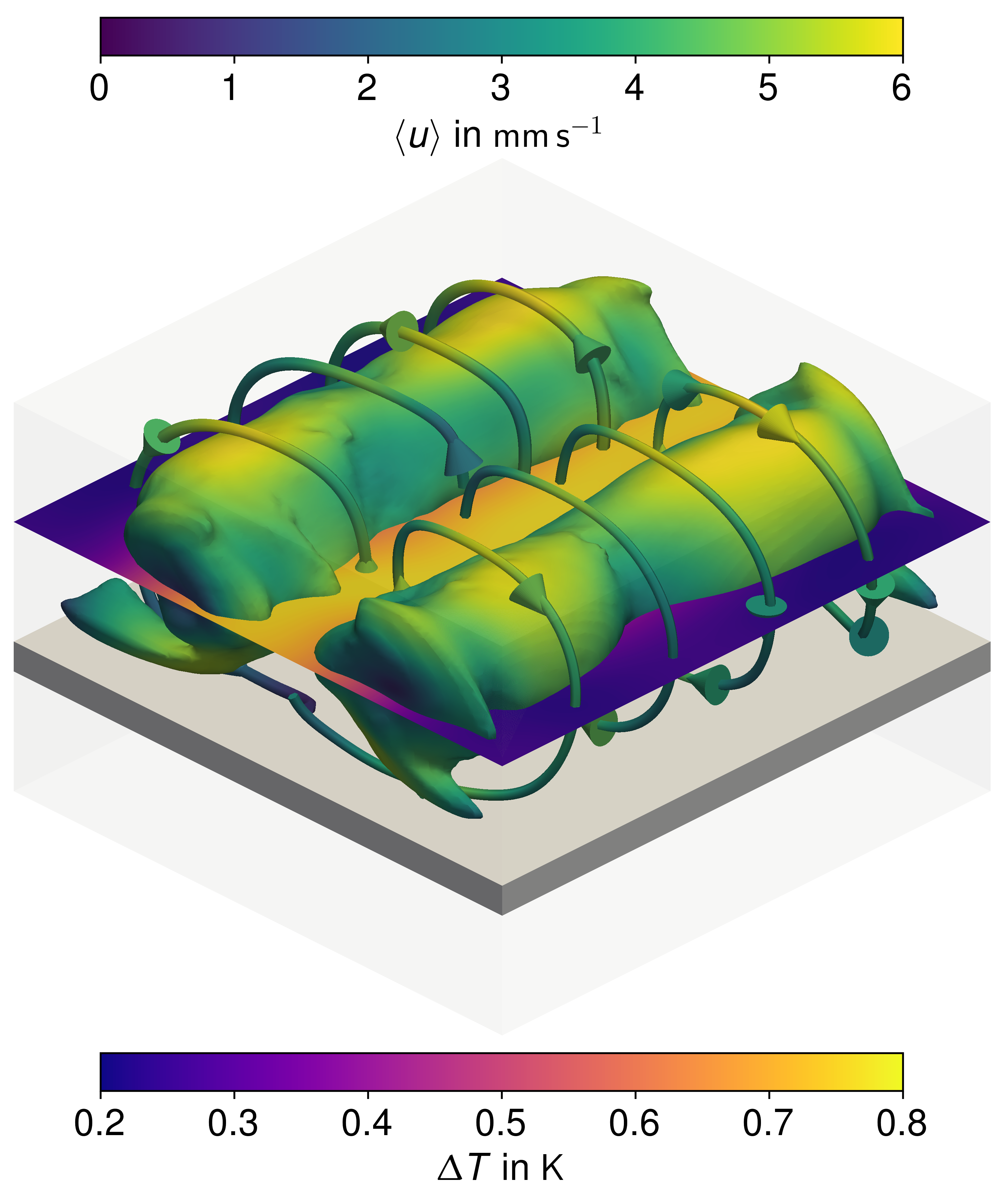}}\hfill
	\subfigure[]{\includegraphics[width=0.5\textwidth]{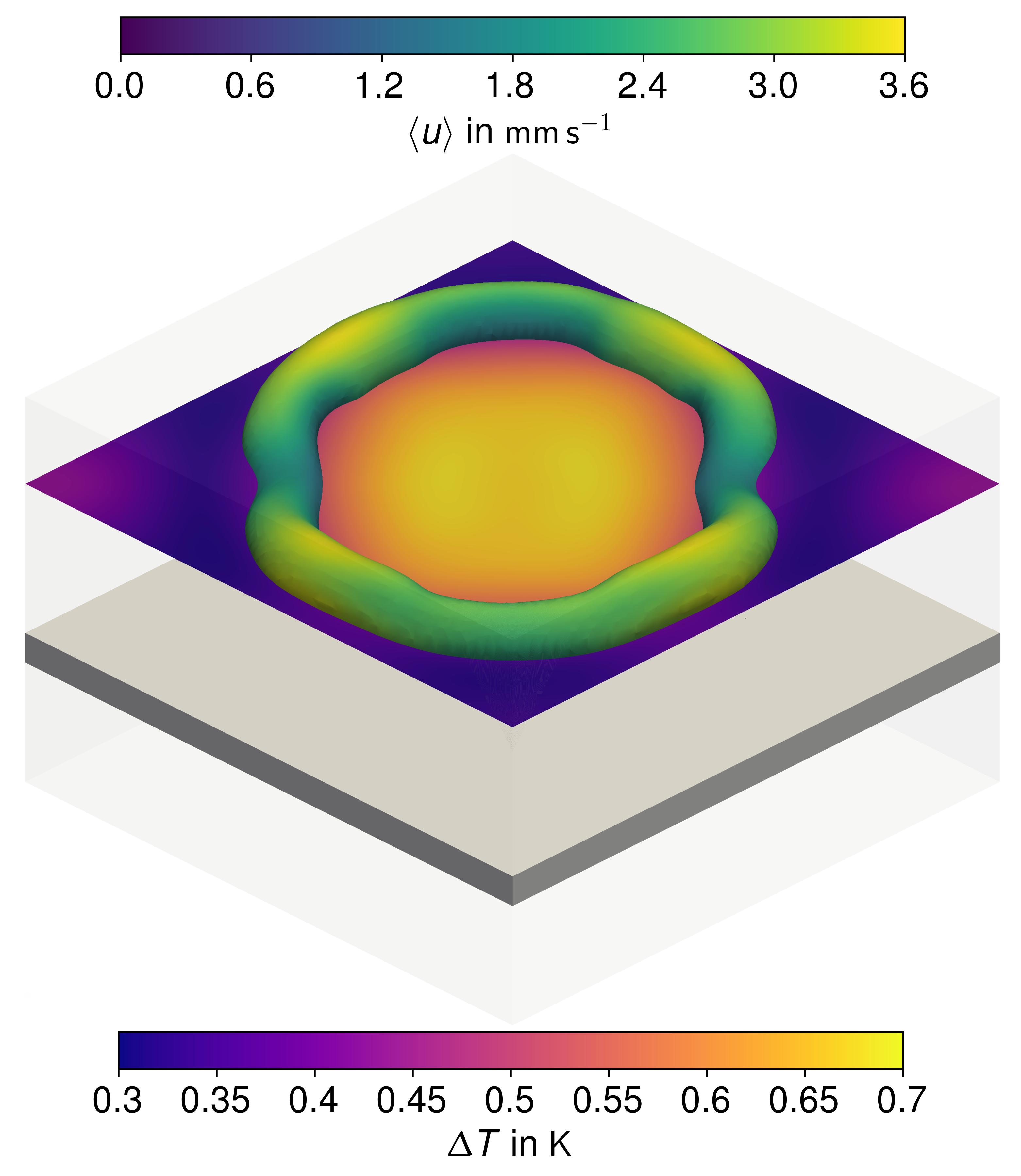}}\hfill
	\caption{Contours of the $\lambda_2$ criterion \cite{Jeong1995},
		flow velocity and temperature distribution in the negative 
		electrode for the case at $x_\text{Li}=\SI{1}{\percent}$ showing two counter rotating vortices
		(a). Flow field and temperature distribution in the negative electrode for 
		the case at $x_\text{Li} = \SI{38}{\percent}$ showing one vortex ring (b). The
		gray layer illustrates the electrolyte.} 
	\label{f:convection3d}
\end{figure}
\begin{figure}[h!]
	\centering
	\subfigure[]{\includegraphics[width=0.5\textwidth]{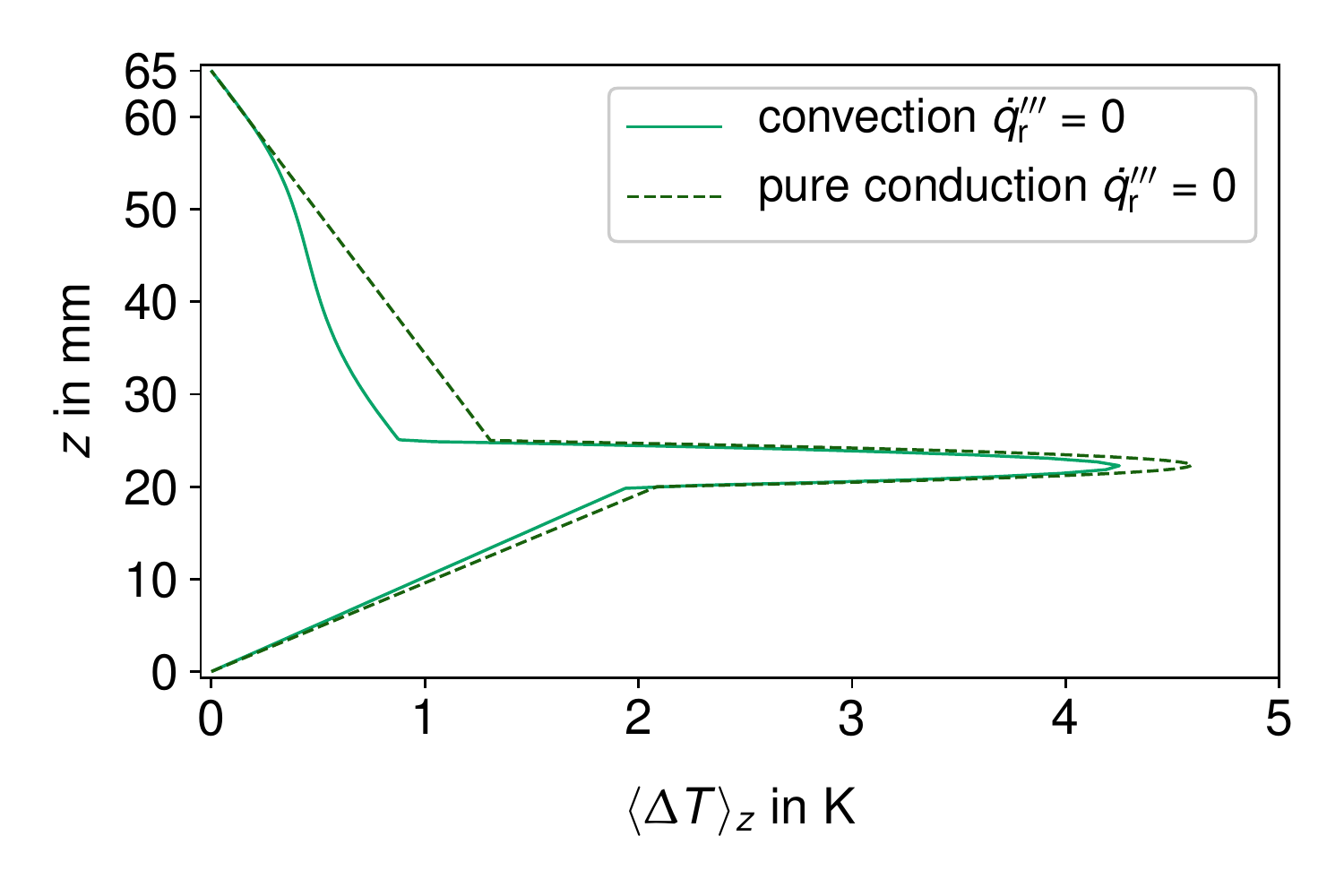}}\hfill
	\subfigure[]{\includegraphics[width=0.5\textwidth]{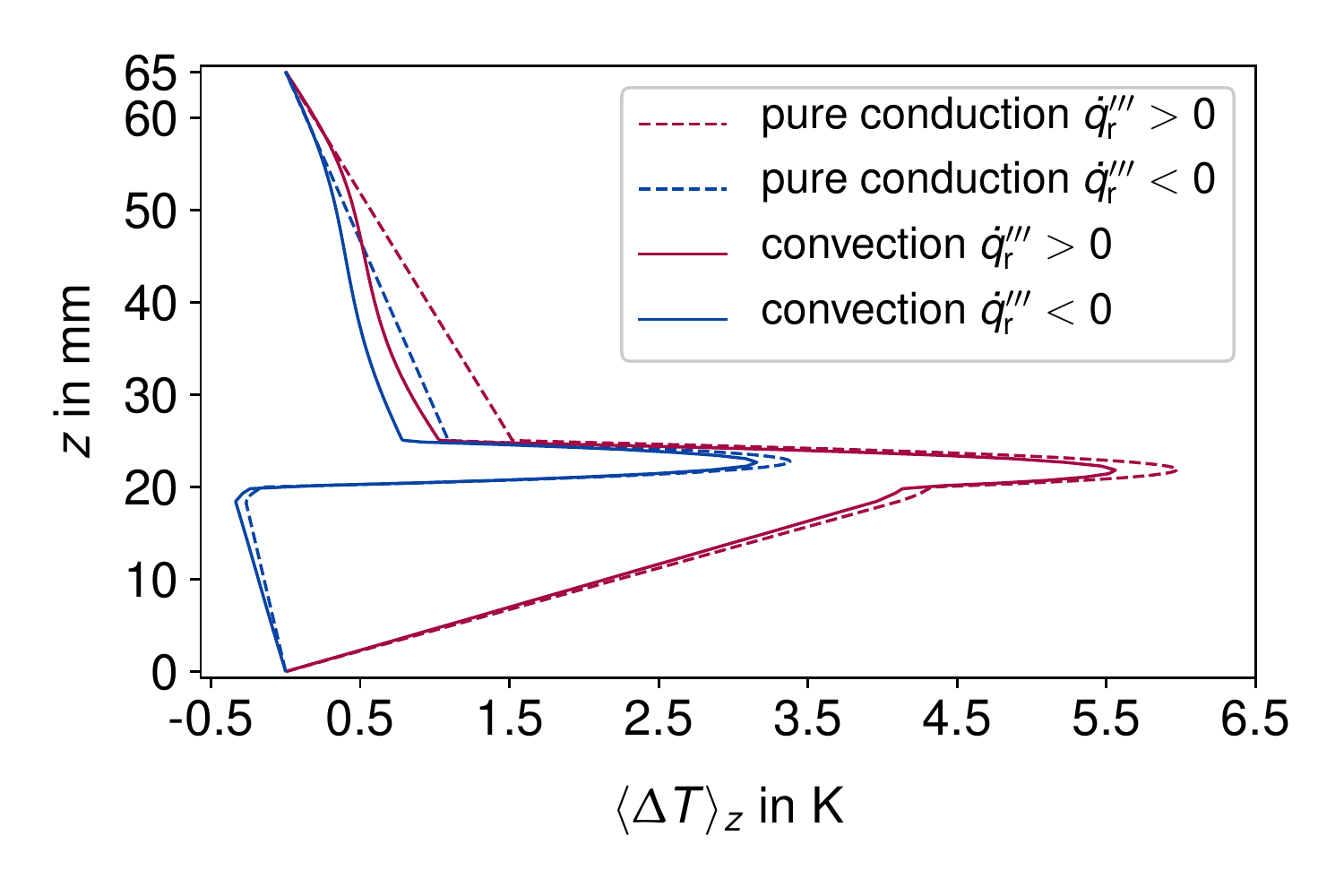}}\hfill
	\subfigure[]{\includegraphics[width=0.5\textwidth]{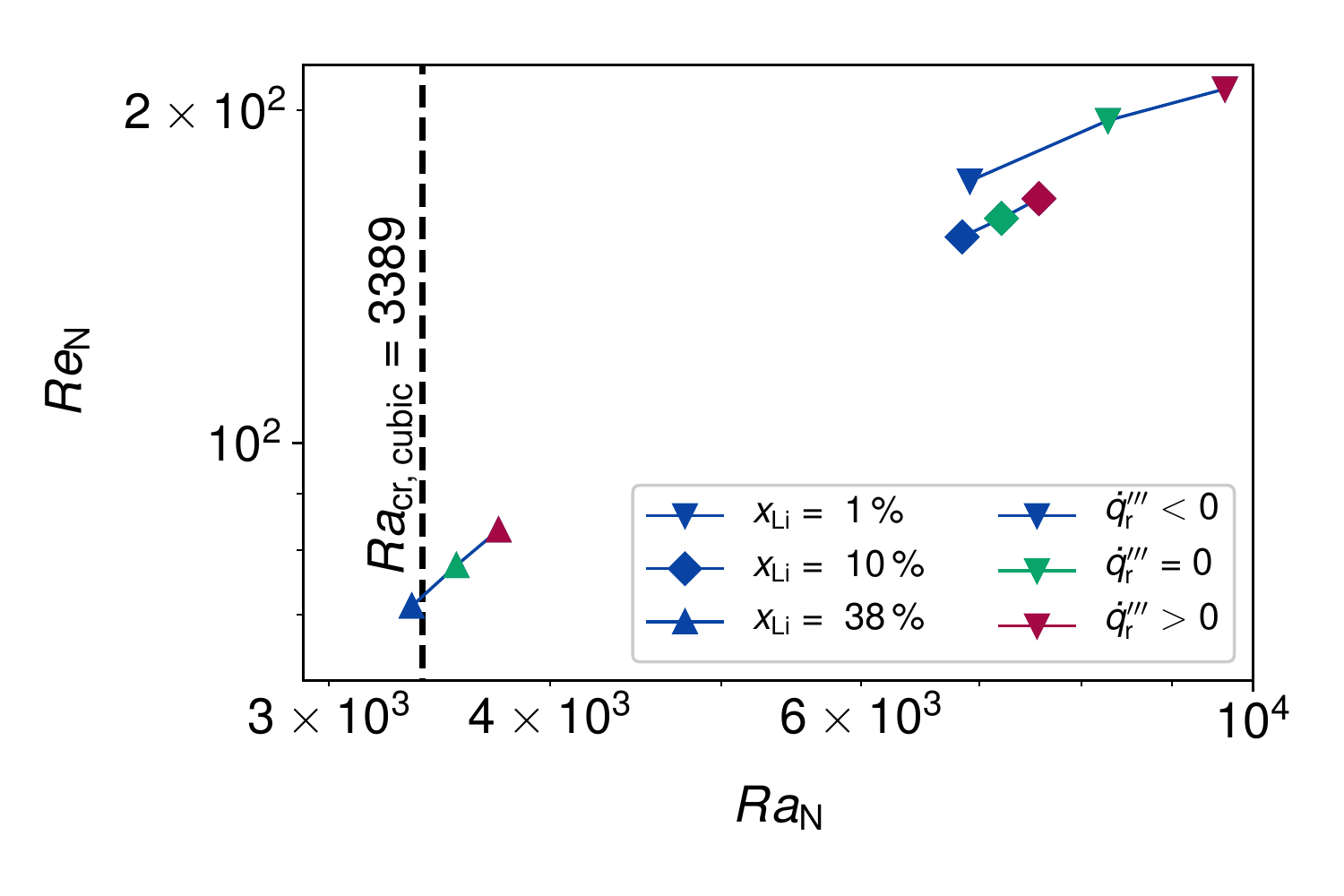}}\hfill
	\subfigure[]{\includegraphics[width=0.5\textwidth]{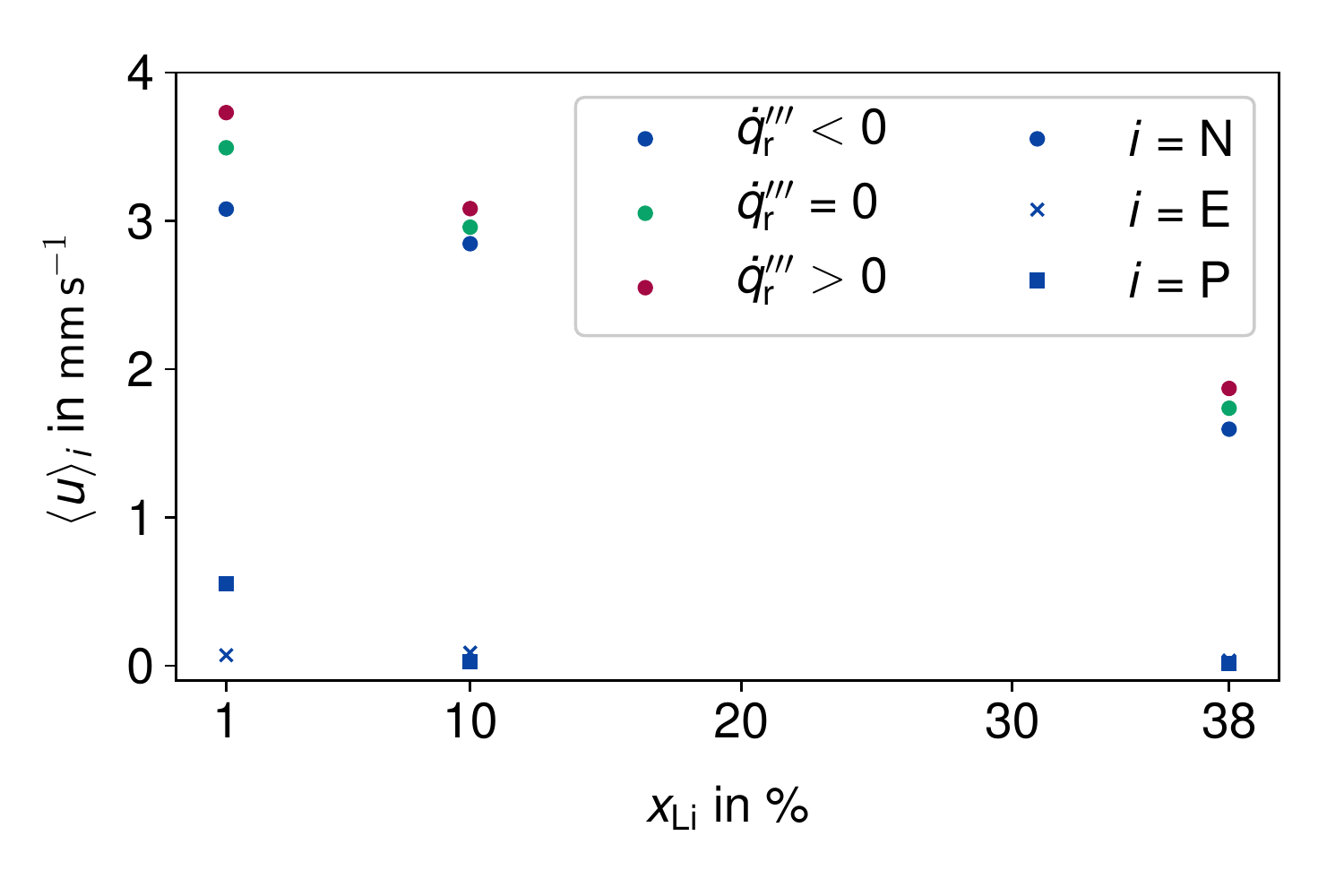}}
	\caption{Three-dimensional simulation with the OpenFOAM solver: 
		Mean temperature along the vertical axis without (a) and
		with electrochemical heat generation (b) for pure conduction and the convection
		case in the fully charged state. Reynolds 
		($Re_\text{N} = \frac{\langle u\rangle_\text{N}\Delta h_\text{N}}{\nu_\text{N}}$) and Rayleigh number 
		($Ra_\text{N} = (\frac{\rho_\text{ref} \cdot c_p \cdot \text{g} \cdot \beta_\text{} \cdot\Delta T_\text{cd} \cdot
			\Delta h_\text{}^3}{\nu_\text{} \cdot k_\text{}})_\text{N}$)
		in the negative electrode  with $\Delta T_\text{cd} $ denoting the conductive temperature difference (c).
		Volume averaged velocity in the three layers depending on the charge state
		with and without electrochemical heat generation (d). The current density is $j =
		\SI{1}{\ampere.\centi\meter^{-2}}$.}
	\label{f:convection}
\end{figure}
The strongest fluid motion is observed in the top electrode, as
shown in Fig.~\ref{f:convection3d}. The flow is driven by
Rayleigh-B\'enard convection, which is caused by a cold
and rigid top wall, and a warm and deformable bottom surface
(RB3 mode of \cite{Goluskin2016}) in a laterally confined box.
Due to the small aspect ratio and the square basis, the flow exhibits
a fully 3D structure with symmetric features \cite{Lappa2005}.
Although we did not find a critical Rayleigh number for the 
exact geometry and boundary conditions of the top layer, eight cases over nine 
exceed the value for a cubic cell ($Ra_\text{cr}= 3389 $)
\cite{Lappa2005,Puigjaner2004} (see Fig.~\ref{f:convection}c).
Fig.~\ref{f:convection3d}a shows the typical flow for the fully
charged LMB: two counter rotating vortices can be observed in the
negative electrode.
During discharge the thickness of the top layer decreases and the aspect ratio
increases from $ AR_\text{N}= \frac{L}{\Delta h_\text{N}} = 2.5 $ to $ AR_\text{N} = 3.14 $. 
The flow structure completely changes and a stable vortex ring appears (Fig.~\ref{f:convection3d}b). 
The presence of a toroidal vortex is a known thermal convection mode 
in this configuration \cite{Gelfgat1999, Pallares1999, Funakoshi2018}. 
The fluid flow reduces the temperature of the side walls compared to
pure conduction. The hot fluid rises in the central part of the cell 
and cools down before descending on lateral walls. \\
In all simulations performed by us, the electrolyte layer remains
in a quiescent state. The unstable density stratification is just too weak
and affects a too small thickness (around \SI{2.5}{\milli\meter}) 
to enhance and sustain a flow. Small velocity perturbations take place
due to the viscous coupling with the negative electrode.
The critical Rayleigh number for a layer subject to an 
internal heat generation and a temperature difference was derived by 
Sparrow et al. \cite{Sparrow1964}. Assuming a laterally infinite layer, rigid
walls and a non-linearity parameter $N_S=\SI{3e3}{}$ we find a
critical Rayleigh number of approximately $Ra_\text{cr} = 583$. This
value is exceeded in one single simulation ($Ra_\text{E} = 671$),
corresponding to a fully charged cell with positive electrochemical 
heat generation. Of course, convection will appear in a thicker
electrolyte -- as shown by previous studies \cite{Shen2015, Koellner2017}. 
Nevertheless, the employment of thin electrolytes in the order of 
\SI{5}{\milli\meter} is important to reduce ohmic losses.

The positive electrode is dominated by conduction. Only in one
single case (fully charged cell and negative electrochemical heat
generation) we obtain a temperature profile with an unstable 
density stratification. Although the Rayleigh number
($Ra_\text{P} = (\frac{\rho_\text{ref} \cdot c_p \cdot \text{g} \cdot \beta_\text{} \cdot \Delta T_\text{cd} \cdot
\Delta h_\text{}^3}{\nu_\text{} \cdot k_\text{}})_\text{P} = 1060$) is smaller than the critical one
($Ra_\text{cr} = 1707$; infinite layer, no slip boundaries),
we observe a weak flow.

Compared to pure conduction, the flow always reduces the observed
temperature (Fig.~\ref{f:convection}a). This effect is especially
strong in the negative electrode, where it lowers the maximum temperature
up to \SI{10}{\percent}. Anyway, the pure conduction model still
provides a good approximation of the vertical temperature
distribution.

The electrochemical heat generation affects the temperature profile in a very
similar way as in pure conduction, as shown in
Fig.~\ref{f:convection}b. While the flow structure does not change
substantially, its magnitude does. Including the electrochemical heat generation
into the model, the flow velocity changes up to 
\SI{12}{\percent} in the negative electrode (Fig.~\ref{f:convection}d).
As already expected \cite{Koellner2017}, no deformation of the
interfaces was observed for all simulated cells.

\section{Summary and outlook}
In this paper we have discussed the thermal phenomena that take place in a
liquid metal battery (LMB) using two different models. We have studied both: 
ohmic and electrochemical heat generation. The 
latter is taken into account for the first time together with thermal convection.

In a first step we developed an analytical 1D conduction model. It provides the vertical
temperature distributions in the cell. This profile, derived from pure conduction,
is the upper bound for the temperature and also the base state over which linear
stability analysis can be performed. It can be used as an initial condition in 
thermo-fluid dynamic simulations.
Furthermore, it allows to identify the region of the cell, which might be affected by
convection and it is a good test model for the evaluation of the importance of
different heat sources. 

The electrochemical heat generation has an impact on the
whole temperature profile. Most importantly, it is able to change the slope of the
temperature profile in the bottom electrode, possibly leading to flow there. 
This very important as it could potentially enhance mass transfer. Compared to pure
ohmic heating, the electrochemical heat is able to
change the maximum cell temperature up to \SI{30}{\percent}.

In a second step, we implemented a fully 3D thermo-fluid dynamic code that allows to study thermal
convection inside the cell. For this purpose, the OpenFOAM Volume of Fluid solver was extended by a
temperature dependent density and the energy equation. The arising problem of
spurious currents was addressed by implementing an additional time step limitation.
Finally, the solver was validated by comparison with a pseudo-spectral code.

Thereafter, the solver was used to study a $10\text{x}\SI{10}{\centi\meter}$ Li$||$Bi 
square cell at three charge states. A mild flow in the order of \si{\milli\meter/\second} 
was observed in the negative electrode. This flow was not able to
deform the interfaces; therefore, we do not expect a short circuit of the
battery to be induced by thermal convection. As our negative electrode consists of
pure Li, the convection there does not affect the mass transport.
However, we expect
that the buoyancy driven flow can be relevant for mass transfer in
multi-element LMBs which use an alloy as top electrode (e.g. \ce{Ca-Mg||Bi}
\cite{Ouchi2016}).

Using a (realistic) electrolyte layer thickness of \SI{5}{\milli\meter} and
a high current density of \SI{1}{\ampere.\centi\meter^{-2}},
we found that the electrolyte is not subject to buoyancy driven flow.
This is not in line with previous studies \cite{Shen2015,Koellner2017}, but 
can be explained by the thin electrolyte layer. Due to viscous coupling (and perhaps
also spurious velocities), only a really weak flow in the order of $\SI{10}{\micro\meter/\second}$
is induced in the electrolyte. 

The stable density stratification in the positive electrode generally 
suppresses any fluid flow. However, it is possible that the electrochemical heat 
promotes convection in the lower electrode, and enhances mixing at
the interface with the electrolyte. We observed that only in one single case.

Finally, we studied three different charge states of the cell. During discharge
not only the location of the temperature peak will shift vertically -- also the
flow structure changes. While we observed two counter rotating vortices in the 
upper electrode of the fully charged cell, one vortex ring appears in the discharged cell
at $x_\text{Li} = \SI{38}{\percent}$. We conclude that
the charge state strongly effects flow field and magnitude.

Our models are strongly simplified. Most importantly, an extended study of
the electrochemical
heat and especially its location should be performed. More detailed modeling should
include radiative heat transfer (in the electrolyte and argon layer) and a 
mass transfer and solidification model. Finally, the boundary conditions will
need improvement: e.g. the insulation and thermal management system may be
included.
Concerning the volume of fluid solver, more work should be dedicated to a 
solution of spurious velocities. The developed model can be applied in the future to
other thermo and electro-metallurgical applications, as e.g. the study of
titanium reduction reactors \cite{Teimurazov2017a,Teimurazov2017}.

\section*{Acknowledgments}
This work was supported by Helmholtz-Gemeinschaft Deutscher
For\-schungs\-zentren (HGF) in frame of the Helmholtz Alliance
``Liquid metal technologies'' (LIMTECH) as well as by the Deutsche
Forschungsgemeinschaft (DFG, German Research Foundation) under award
number 338560565 and grant number KO 5515/1-1. The computations were
performed on the Bull HPC-Cluster ``Taurus'' at the Center for
Information Services and High Performance Computing (ZIH) at TU
Dresden, on the cluster ``Hydra'' at Helmholtz-Zentrum Dresden --
Rossendorf and at the Computing Center of TU Ilmenau. Fruitful
discussions with V. Galindo, D. Kelley, A. Teimurazov, V. Vuk\v
cevi\'c and O. Zikanov on several aspects of thermal convection
and liquid metal batteries are gratefully acknowledged. N. Weber
thanks H. Schulz for the HPC support.

\section*{References}
\bibliographystyle{elsarticle-num}
\bibliography{literatur}

\begin{thebibliography}{100}
\expandafter\ifx\csname url\endcsname\relax
  \def\url#1{\texttt{#1}}\fi
\expandafter\ifx\csname urlprefix\endcsname\relax\def\urlprefix{URL }\fi
\expandafter\ifx\csname href\endcsname\relax
  \def\href#1#2{#2} \def\path#1{#1}\fi

\bibitem{Lindley:2010}
D.~Lindley, Smart grids: The energy storage problem, Nature 463~(7277) (2010)
  18--20.

\bibitem{Pickard:2015}
W.~F. Pickard, Massive electricity storage for a developed economy of ten
  billion people, IEEE Access 3 (2015) 1392--1407.

\bibitem{Spatocco2015}
B.~L. Spatocco, D.~R. Sadoway, Cost-{{Based Discovery}} for {{Engineering
  Solutions}}, in: R.~C. Alkire, P.~N. Bartlett, J.~Lipkowski (Eds.),
  Electrochemical {{Engineering Across Scales}}: {{From Molecules}} to
  {{Processes}}, Vol.~15 of Advances in Electrochemical Science and
  Engineering, {Wiley-VCH}, 2015, pp. 227--262.

\bibitem{Kim2013b}
H.~Kim, D.~A. Boysen, J.~M. Newhouse, B.~L. Spatocco, B.~Chung, P.~J. Burke,
  D.~J. Bradwell, K.~Jiang, A.~A. Tomaszowska, K.~Wang, W.~Wei, L.~A. Ortiz,
  S.~A. Barriga, S.~M. Poizeau, D.~R. Sadoway, Liquid {{Metal Batteries}}:
  {{Past}}, {{Present}}, and {{Future}}, Chem. Rev. 113~(3) (2013) 2075--2099.
\newblock \href {http://dx.doi.org/10.1021/cr300205k}
  {\path{doi:10.1021/cr300205k}}.

\bibitem{Agruss1963a}
B.~Agruss, Nuclear {{Liquid Metal Cell For Space Power}}, in: Proceedings of
  the 17th Power Sources Conference, 1963, pp. 100--103.

\bibitem{Cairns1967}
E.~J. Cairns, C.~E. Crouthamel, A.~K. Fischer, M.~S. Foster, J.~C. Hesson,
  C.~E. Johnson, H.~Shimotake, A.~D. Tevebaugh, Galvanic {{Cells}} with
  {{Fused}}-{{Salt Electrolytes}}, ANL-7316, {Argonne National Laboratory},
  1967.

\bibitem{Cairns1969b}
E.~J. Cairns, H.~Shimotake, High-{{Temperature Batteries}}, Science 164 (1969)
  1347--1355.

\bibitem{Kim2013a}
H.~Kim, D.~A. Boysen, T.~Ouchi, D.~R. Sadoway, Calcium - bismuth electrodes for
  large - scale energy storage, J. Power Sources 241 (2013) 239--248.

\bibitem{Kelley2017}
D.~H. Kelley, T.~Weier, Fluid mechanics of liquid metal batteries, Appl. Mech.
  Rev. 70~(2) (2018) 020801.
\newblock \href {http://dx.doi.org/10.1115/1.4038699}
  {\path{doi:10.1115/1.4038699}}.

\bibitem{Ning2015}
X.~Ning, S.~Phadke, B.~Chung, H.~Yin, P.~Burke, D.~R. Sadoway, Self-healing
  {{Li}}\textendash{}{{Bi}} liquid metal battery for grid-scale energy storage,
  J. Power Sources 275 (2015) 370--376.
\newblock \href {http://dx.doi.org/10.1016/j.jpowsour.2014.10.173}
  {\path{doi:10.1016/j.jpowsour.2014.10.173}}.

\bibitem{Stefani2011}
F.~Stefani, T.~Weier, T.~Gundrum, G.~Gerbeth, How to circumvent the size
  limitation of liquid metal batteries due to the {{Tayler}} instability,
  Energy Convers. Manag. 52 (2011) 2982--2986.
\newblock \href {http://dx.doi.org/10.1016/j.enconman.2011.03.003}
  {\path{doi:10.1016/j.enconman.2011.03.003}}.

\bibitem{Weber2013}
N.~Weber, V.~Galindo, F.~Stefani, T.~Weier, T.~Wondrak, Numerical simulation of
  the {{Tayler}} instability in liquid metals, New J. Phys. 15 (2013) 043034.
\newblock \href {http://dx.doi.org/10.1088/1367-2630/15/4/043034}
  {\path{doi:10.1088/1367-2630/15/4/043034}}.

\bibitem{Weber2014}
N.~Weber, V.~Galindo, F.~Stefani, T.~Weier, Current-driven flow instabilities
  in large-scale liquid metal batteries, and how to tame them, J. Power Sources
  265 (2014) 166--173.
\newblock \href {http://dx.doi.org/10.1016/j.jpowsour.2014.03.055}
  {\path{doi:10.1016/j.jpowsour.2014.03.055}}.

\bibitem{Herreman2015}
W.~Herreman, C.~Nore, L.~Cappanera, J.-L. Guermond, Tayler instability in
  liquid metal columns and liquid metal batteries, J. Fluid Mech. 771 (2015)
  79--114.
\newblock \href {http://dx.doi.org/10.1017/jfm.2015.159}
  {\path{doi:10.1017/jfm.2015.159}}.

\bibitem{Weber2015b}
N.~Weber, V.~Galindo, F.~Stefani, T.~Weier, The {{Tayler}} instability at low
  magnetic {{Prandtl}} numbers: Between chiral symmetry breaking and helicity
  oscillations, New J. Phys. 17~(11) (2015) 113013.
\newblock \href {http://dx.doi.org/10.1088/1367-2630/17/11/113013}
  {\path{doi:10.1088/1367-2630/17/11/113013}}.

\bibitem{Weier2017}
T.~Weier, A.~Bund, W.~El-Mofid, G.~M. Horstmann, C.-C. Lalau, S.~Landgraf,
  M.~Nimtz, M.~Starace, F.~Stefani, N.~Weber, Liquid metal batteries -
  materials selection and fluid dynamics, IOP Conf. Ser. Mater. Sci. Eng. 228
  (2017) 012013.
\newblock \href {http://dx.doi.org/10.1088/1757-899X/228/1/012013}
  {\path{doi:10.1088/1757-899X/228/1/012013}}.

\bibitem{Bradwell2015}
D.~Bradwell, G.~Ceder, L.~A. Ortiz, D.~R. Sadoway, Liquid metal alloy energy
  storage device, {US} patent 9,076,996 B2 (2015).

\bibitem{Weber2014b}
N.~Weber, V.~Galindo, J.~Priede, F.~Stefani, T.~Weier, The influence of current
  collectors on {{Tayler}} instability and electro vortex flows in liquid metal
  batteries, Phys. Fluids 27 (2015) 014103.
\newblock \href {http://dx.doi.org/10.1063/1.4905325}
  {\path{doi:10.1063/1.4905325}}.

\bibitem{Stefani2015}
F.~Stefani, V.~Galindo, C.~Kasprzyk, S.~Landgraf, M.~Seilmayer, M.~Starace,
  N.~Weber, T.~Weier, Magnetohydrodynamic effects in liquid metal batteries,
  IOP Conf. Ser. Mater. Sci. Eng. 143 (2016) 012024.
\newblock \href {http://dx.doi.org/10.1088/1757-899X/143/1/012024}
  {\path{doi:10.1088/1757-899X/143/1/012024}}.

\bibitem{Ashour2017a}
R.~Ashour, D.~H. Kelley, A.~Salas, M.~Starace, N.~Weber, T.~Weier, Competing
  forces in liquid metal electrodes and batteries, J. Power Sources 378 (2018)
  301--310.
\newblock \href {http://dx.doi.org/10.1016/j.jpowsour.2017.12.042}
  {\path{doi:10.1016/j.jpowsour.2017.12.042}}.

\bibitem{Weber2018}
N.~Weber, M.~Nimtz, P.~Personnettaz, A.~Salas, T.~Weier, Electromagnetically
  driven convection suitable for mass transfer enhancement in liquid metal
  batteries, arXiv:1802.02214.

\bibitem{Zikanov2015}
O.~Zikanov, Metal pad instabilities in liquid metal batteries, Phys. Rev. E 92
  (2015) 063021.

\bibitem{Weber2017}
N.~Weber, P.~Beckstein, V.~Galindo, W.~Herreman, C.~Nore, F.~Stefani, T.~Weier,
  Metal pad roll instability in liquid metal batteries, Magnetohydrodynamics
  53~(1) (2017) 129--140.

\bibitem{Weber2017a}
N.~Weber, P.~Beckstein, W.~Herreman, G.~M. Horstmann, C.~Nore, F.~Stefani,
  T.~Weier, Sloshing instability and electrolyte layer rupture in liquid metal
  batteries, Phys. Fluids 29~(5) (2017) 054101.
\newblock \href {http://dx.doi.org/10.1063/1.4982900}
  {\path{doi:10.1063/1.4982900}}.

\bibitem{Bojarevics2017}
V.~Bojarevics, A.~Tucs, {{MHD}} of {{Large Scale Liquid Metal Batteries}}, in:
  A.~P. Ratvik (Ed.), Light {{Metals}} 2017, {Springer International
  Publishing}, Cham, 2017, pp. 687--692.

\bibitem{Horstmann2017}
G.~M. Horstmann, N.~Weber, T.~Weier, Coupling and stability of interfacial
  waves in liquid metal batteries, J. Fluid Mech. 845 (2018) 1--35.
\newblock \href {http://dx.doi.org/10.1017/jfm.2018.223}
  {\path{doi:10.1017/jfm.2018.223}}.

\bibitem{Zikanov2017}
O.~Zikanov, Shallow water modeling of rolling pad instability in liquid metal
  batteries, arXiv:1706.08589.

\bibitem{Wang2015}
W.~Wang, K.~Wang, Simulation of thermal properties of the liquid metal
  batteries, in: 6th {{International Conference}} on {{Power Electronics
  Systems}} and {{Applications}} ({{PESA}}), {IEEE}, 2015, pp. 1--11.

\bibitem{Kelley2014}
D.~H. Kelley, D.~R. Sadoway, Mixing in a liquid metal electrode, Phys. Fluids
  26~(5) (2014) 057102.
\newblock \href {http://dx.doi.org/10.1063/1.4875815}
  {\path{doi:10.1063/1.4875815}}.

\bibitem{Perez2015}
A.~Perez, D.~H. Kelley, Ultrasound {{Velocity Measurement}} in a {{Liquid Metal
  Electrode}}, J. Vis. Exp. 102 (2015) e52622.
\newblock \href {http://dx.doi.org/10.3791/52622} {\path{doi:10.3791/52622}}.

\bibitem{Beltran2016}
A.~Beltr{\'a}n, {{MHD}} natural convection flow in a liquid metal electrode,
  Appl. Therm. Eng. 114 (2016) 1203--1212.
\newblock \href {http://dx.doi.org/10.1016/j.applthermaleng.2016.09.006}
  {\path{doi:10.1016/j.applthermaleng.2016.09.006}}.

\bibitem{Shen2015}
Y.~Shen, O.~Zikanov, Thermal convection in a liquid metal battery, Theor.
  Comput. Fluid Dyn. 30~(4) (2016) 275--294.
\newblock \href {http://dx.doi.org/10.1007/s00162-015-0378-1}
  {\path{doi:10.1007/s00162-015-0378-1}}.

\bibitem{Koellner2017}
T.~K{\"o}llner, T.~Boeck, J.~Schumacher, Thermal {{Rayleigh}}-{{Marangoni}}
  convection in a three-layer liquid-metal-battery model, Phys. Rev. E 95
  (2017) 053114.
\newblock \href {http://dx.doi.org/10.1103/PhysRevE.95.053114}
  {\path{doi:10.1103/PhysRevE.95.053114}}.

\bibitem{Newman2004}
J.~Newman, K.~E. Thomas-Alyea, Electrochemical {{Systems}}, {John Wiley \&
  Sons}, 2004.

\bibitem{Heines1972}
H.~Heines, J.~Westwater, The effect of heats of solution on marangoni
  convection, Int. J. Heat Mass Transf. 15~(11) (1972) 2109--2117.
\newblock \href {http://dx.doi.org/10.1016/0017-9310(72)90035-X}
  {\path{doi:10.1016/0017-9310(72)90035-X}}.

\bibitem{Eckert1999}
K.~Eckert, A.~Grahn, Plume and finger regimes driven by an exothermic
  interfacial reaction, Phys. Rev. Lett. 82~(22) (1999) 4436.

\bibitem{Grube1934}
H.~Grube, G.~Vosskuehler, S.~H., Elektrische {L}eitf{\"a}higkeit und
  {Z}ustandsdiagramm bei bin{\"a}ren {L}egierungen. {D}as {S}ystem
  {L}ithium-{W}ismuth., Z. Elektrochem. angew. p. 40~(5) (1934) 270--274.

\bibitem{Zintl1935}
E.~Zintl, G.~Brauer, Metals and alloys. 14. constitution of {B}i-{L}i alloys,
  Z. Elektochem. 41 (1935) 297--303.

\bibitem{Seith1937}
W.~Seith, O.~Kubaschewski, The heats of formation of several alloys, Z.
  Elektrochem. 43 (1937) 743--749.

\bibitem{Shchukarev1957}
S.~Shchukarev, M.~Morozova, K.~Yn, V.~Sharov, Enthalpy of formation of {Li} and
  {Ba} bismuthides, Zh. Obshchei Khim 27(2) (1957) 290--93.

\bibitem{Hansen1958}
M.~Hansen, Constitution of Binary Alloys, {McGraw-Hill}, 1958.

\bibitem{Foster1964}
M.~S. Foster, S.~E. Wood, C.~E. Crouthamel, Thermodynamics of {{Binary
  Alloys}}. {{I}}. {{The Lithium}}-{{Bismuth System}}, Inorg. Chem. 3~(10)
  (1964) 1428--1431.

\bibitem{Demidov1973}
A.~Demidov, A.~G. Morachevskii, Thermodynamic properties of liquid lithium -
  bismuth alloys, Elektrokhimiya 9~(9) (1973) 1393--1394.

\bibitem{Saboungi1978}
M.-L. Saboungi, J.~Marr, M.~Blander, Thermodynamic properties of a quasi-ionic
  alloy from electromotive force measurements: {{The Li}}\textendash{}{{Pb}}
  system, J. Chem. Phys. 68~(4) (1978) 1375.
\newblock \href {http://dx.doi.org/10.1063/1.435957}
  {\path{doi:10.1063/1.435957}}.

\bibitem{Predel1979}
B.~Predel, G.~Oehme, Calorimetric investigation of liquid {Li}--{Tl},
  {Li}--{In}, and {Li}--{Bi} alloys, Z. Metallkd. 70~(9) (1979) 618--623.

\bibitem{Gasior1994}
W.~Gasior, Z.~Moser, W.~Zakulski, Bi-{{Li System}}. {{Thermodynamic}}
  properties and the phase diagram calculations, Arch. Metall. 39~(4) (1994)
  355--364.

\bibitem{Liu2013}
J.~Liu, J.-G. Zhang, Z.~Yang, J.~P. Lemmon, C.~Imhoff, G.~L. Graff, L.~Li,
  J.~Hu, C.~Wang, J.~Xiao, G.~Xia, V.~V. Viswanathan, S.~Baskaran, V.~Sprenkle,
  X.~Li, Y.~Shao, B.~Schwenzer, Materials {{Science}} and {{Materials
  Chemistry}} for {{Large Scale Electrochemical Energy Storage}}: {{From
  Transportation}} to {{Electrical Grid}}, Adv. Funct. Mater. 23~(8) (2013)
  929--946.
\newblock \href {http://dx.doi.org/10.1002/adfm.201200690}
  {\path{doi:10.1002/adfm.201200690}}.

\bibitem{Cao2014}
Z.~Cao, W.~Xie, P.~Chartrand, S.~Wei, G.~Du, Z.~Qiao, Thermodynamic assessment
  of the {{Bi}}-alkali metal ({{Li}}, {{Na}}, {{K}}, {{Rb}}) systems using the
  modified quasichemical model for the liquid phase, Calphad 46 (2014)
  159--167.
\newblock \href {http://dx.doi.org/10.1016/j.calphad.2014.04.001}
  {\path{doi:10.1016/j.calphad.2014.04.001}}.

\bibitem{Bernardi1985}
D.~Bernardi, E.~Pawlikowski, J.~Newman, A general energy balance for battery
  systems, J. Electrochem. Soc. 132~(1) (1985) 5--12.

\bibitem{Braunstein1971}
J.~Braunstein, G.~Mamantov, G.~P. Smith (Eds.), Advances in {{Molten Salt
  Chemistry}}, {Springer US}, Boston, MA, 1971.

\bibitem{Newhouse2017}
J.~M. Newhouse, D.~R. Sadoway, Charge-{{Transfer Kinetics}} of {{Alloying}} in
  {{Mg}}-{{Sb}} and {{Li}}-{{Bi Liquid Metal Electrodes}}, J. Electrochem. Soc.
  164~(12) (2017) A2665--A2669.

\bibitem{Shimotake1967}
H.~Shimotake, E.~J. Cairns, Bimetallic galvanic cells with fused-salt
  electrolytes, in: Advances in {{Energy Conversion Engineering}}, {ASME},
  Florida, 1967, pp. 951--962.

\bibitem{Chum1981}
H.~L. Chum, R.~A. Osteryoung, Review of Thermally Regenerative Electrochemical
  Cells, {Solar Energy Research Institute}, 1981.

\bibitem{Bradwell2016b}
D.~J. Bradwell, H.~Nayar, Z.~T. Modest, S.~L. Golmon, Thermal management of
  liquid metal batteries, {US} patent US20160365612 A1 (2016).

\bibitem{Carslaw1959}
H.~S. Carslaw, J.~C. Jaeger, Conduction of {{Heat}} in {{Solids}}, {Clarendon
  Press}, 1959.

\bibitem{Gu2000}
W.~B. Gu, C.~Y. Wang, Thermal-electrochemical modeling of battery systems, J.
  Electrochem. Soc. 147~(8) (2000) 2910--2922.

\bibitem{Kumaresan2008}
K.~Kumaresan, G.~Sikha, R.~E. White, Thermal model for a {Li}-ion cell, J.
  Electrochem. Soc. 155~(2) (2008) A164--A171.

\bibitem{Min2012}
J.~K. Min, C.-H. Lee, Numerical study on the thermal management system of a
  molten sodium-sulfur battery module, J. Power Sources 210 (2012) 101--109.

\bibitem{Deen1998}
W.~M. Deen, Analysis of Transport Phenomena, Topics in chemical engineering,
  {Oxford University Press}, New York, 1998.

\bibitem{Personnettaz2017}
P.~Personnettaz, Assessment of thermal phenomena in {{Li}}$||${{Bi}} liquid
  metal batteries through analytical and numerical models, master thesis,
  Politecnico di Torino (2017).

\bibitem{Worner2003}
M.~W{\"o}rner, A compact introduction to the numerical modeling of multiphase
  flows, Forschungszentrum Karlsruhe, 2003.

\bibitem{Weller1998}
H.~G. Weller, G.~Tabor, H.~Jasak, C.~Fureby, A tensorial approach to
  computational continuum mechanics using object-oriented techniques, Comput.
  Phys. 12~(6) (1998) 620--631.

\bibitem{Rusche2002}
H.~Rusche, Computational {{Fluid Dynamics}} of {{Dispersed Two}}-{{Phase
  Flows}} at {{High Phase Fractions}}, Ph.D. thesis, Imperial College London
  (2002).

\bibitem{Ubbink1997}
O.~Ubbink, Numerical prediction of two fluid systems with sharp interfaces,
  Ph.D. thesis, University of London (1997).

\bibitem{Brackbill1992}
J.~U. Brackbill, D.~B. Kothe, C.~Zemach, A continuum method for modeling
  surface tension, J. Comput. Phys. 100 (1992) 335--354.

\bibitem{Kissling2010}
K.~Kissling, J.~Springer, H.~Jasak, S.~Schutz, K.~Urban, M.~Piesche, A coupled
  pressure based solution algorithm based on the volume-of-fluid approach for
  two or more immiscible fluids, in: Proceedings of the V European Conference
  on Computational Fluid Dynamics ECCOMAS CFD 2010, 2010, pp. 1--16.

\bibitem{Carson2005}
J.~K. Carson, S.~J. Lovatt, D.~J. Tanner, A.~C. Cleland, Thermal conductivity
  bounds for isotropic, porous materials, Int. J. Heat Mass Transf. 48~(11)
  (2005) 2150--2158.
\newblock \href {http://dx.doi.org/10.1016/j.ijheatmasstransfer.2004.12.032}
  {\path{doi:10.1016/j.ijheatmasstransfer.2004.12.032}}.

\bibitem{Kumar2014}
S.~S. Kumar, Y.~M.~C. Delaur{\'e}, Convective {{Heat Transfer}} of an {{Air
  Bubble}} in {{Water With Variable Thermophysical Properties}}, Heat Transf.
  Eng. 35~(14-15) (2014) 1370--1379.
\newblock \href {http://dx.doi.org/10.1080/01457632.2013.877317}
  {\path{doi:10.1080/01457632.2013.877317}}.

\bibitem{Nabil2016}
M.~Nabil, A.~S. Rattner, {{interThermalPhaseChangeFoam}}-{{A}} framework for
  two-phase flow simulations with thermally driven phase change, SoftwareX 5
  (2016) 216--226.
\newblock \href {http://dx.doi.org/10.1016/j.softx.2016.10.002}
  {\path{doi:10.1016/j.softx.2016.10.002}}.

\bibitem{Greenshields2016}
C.~Greenshields, {{OpenFOAM User Guide}} (2016).

\bibitem{Boeck2002}
T.~Boeck, A.~Nepomnyashchy, I.~Simanovskii, A.~Golovin, L.~Braverman, A.~Thess,
  Three-dimensional convection in a two-layer system with anomalous
  thermocapillary effect, Phys. Fluids 14~(11) (2002) 3899--3911.
\newblock \href {http://dx.doi.org/10.1063/1.1506923}
  {\path{doi:10.1063/1.1506923}}.

\bibitem{Boeck2003}
T.~Boeck, M.~Jurgk, U.~Bahr, Oscillatory {{Rayleigh}}-{{Marangoni}} convection
  in a layer heated from above: {{Numerical}} simulations with an undeformable
  free surface, Phys. Rev. E 67~(2) (2003) 027303.
\newblock \href {http://dx.doi.org/10.1103/PhysRevE.67.027303}
  {\path{doi:10.1103/PhysRevE.67.027303}}.

\bibitem{Koellner2014}
T.~K{\"o}llner, M.~Rossi, F.~Broer, T.~Boeck, Chemical convection in the
  methylene-blue\textendash{}glucose system: {{Optimal}} perturbations and
  three-dimensional simulations, Phys. Rev. E 90~(5) (2014) 053004.
\newblock \href {http://dx.doi.org/10.1103/PhysRevE.90.053004}
  {\path{doi:10.1103/PhysRevE.90.053004}}.

\bibitem{Koellner2016}
T.~K{\"o}llner, K.~Schwarzenberger, K.~Eckert, T.~Boeck, The eruptive regime of
  mass-transfer-driven {{Rayleigh}}\textendash{}{{Marangoni}} convection, J.
  Fluid Mech. 791 (2016) R4.
\newblock \href {http://dx.doi.org/10.1017/jfm.2016.63}
  {\path{doi:10.1017/jfm.2016.63}}.

\bibitem{Jeong1995}
J.~Jeong, F.~Hussain, On the identification of a vortex, Journal of fluid
  mechanics 285 (1995) 69--94.

\bibitem{Goluskin2016}
D.~Goluskin, Internally {{Heated Convection}} and {{Rayleigh}}-{{B{\'e}nard
  Convection}}, {Springer International Publishing}, Cham, 2016.
\newblock \href {http://dx.doi.org/10.1007/978-3-319-23941-5}
  {\path{doi:10.1007/978-3-319-23941-5}}.

\bibitem{Lappa2005}
M.~Lappa, On the nature and structure of possible three-dimensional steady
  flows in closed and open parallelepipedic and cubical containers under
  different heating conditions and driving forces, Fluid Mechanics and Material
  Processing 1 (2005) 1--19.

\bibitem{Puigjaner2004}
D.~Puigjaner, J.~Herrero, F.~Giralt, C.~Sim{\'o}, Stability analysis of the
  flow in a cubical cavity heated from below, Phys. Fluids 16~(10) (2004)
  3639--3655.

\bibitem{Gelfgat1999}
A.~Y. Gelfgat, Different {{Modes}} of {{Rayleigh}}-{{B{\'e}nard Instability}}
  in {{Two}}- and {{Three}}-{{Dimensional Rectangular Enclosures}}, J Comput
  Phys 156 (1999) 300 -- 324.

\bibitem{Pallares1999}
J.~Pallares, F.~Grau, F.~Giralt, Flow transitions in laminar
  rayleigh--b{\'e}nard convection in a cubical cavity at moderate rayleigh
  numbers, International Journal of Heat and Mass Transfer 42~(4) (1999)
  753--769.

\bibitem{Funakoshi2018}
M.~Funakoshi, Onset of thermal convection in a rectangular parallelepiped
  cavity of small aspect ratios, Fluid Dynamics Research 50~(2) (2018) 021402.

\bibitem{Sparrow1964}
E.~Sparrow, R.~Goldstein, V.~Jonsson, Thermal instability in a horizontal fluid
  layer: effect of boundary conditions and non-linear temperature profile, J.
  Fluid Mech. 18~(4) (1964) 513--528.

\bibitem{Ouchi2016}
T.~Ouchi, H.~Kim, B.~L. Spatocco, D.~R. Sadoway, Calcium-based multi-element
  chemistry for grid-scale electrochemical energy storage, Nat. Commun. 7
  (2016) 10999.
\newblock \href {http://dx.doi.org/10.1038/ncomms10999}
  {\path{doi:10.1038/ncomms10999}}.

\bibitem{Teimurazov2017a}
A.~Teimurazov, P.~Frick, F.~Stefani, Thermal convection of liquid metal in the
  titanium reduction reactor, IOP Conf. Ser. Mater. Sci. Eng. 208 (2017)
  012041.
\newblock \href {http://dx.doi.org/10.1088/1757-899X/208/1/012041}
  {\path{doi:10.1088/1757-899X/208/1/012041}}.

\bibitem{Teimurazov2017}
A.~Teimurazov, P.~Frick, N.~Weber, F.~Stefani, Numerical simulations of
  convection in the titanium reduction reactor, J. Phys. Conf. Ser. 891 (2017)
  012076.
\newblock \href {http://dx.doi.org/10.1088/1742-6596/891/1/012076}
  {\path{doi:10.1088/1742-6596/891/1/012076}}.

\bibitem{Raeini2012}
A.~Q. Raeini, M.~J. Blunt, B.~Bijeljic, Modelling two-phase flow in porous
  media at the pore scale using the volume-of-fluid method, J. Comput. Phys.
  231~(17) (2012) 5653--5668.
\newblock \href {http://dx.doi.org/10.1016/j.jcp.2012.04.011}
  {\path{doi:10.1016/j.jcp.2012.04.011}}.

\bibitem{Harvie2006}
D.~Harvie, M.~Davidson, M.~Rudman, An analysis of parasitic current generation
  in {{Volume}} of {{Fluid}} simulations, Appl. Math. Model. 30~(10) (2006)
  1056--1066.
\newblock \href {http://dx.doi.org/10.1016/j.apm.2005.08.015}
  {\path{doi:10.1016/j.apm.2005.08.015}}.

\bibitem{Abadie2015}
T.~Abadie, J.~Aubin, D.~Legendre, On the combined effects of surface tension
  force calculation and interface advection on spurious currents within
  {{Volume}} of {{Fluid}} and {{Level Set}} frameworks, J. Comput. Phys. 297
  (2015) 611--636.
\newblock \href {http://dx.doi.org/10.1016/j.jcp.2015.04.054}
  {\path{doi:10.1016/j.jcp.2015.04.054}}.

\bibitem{Vukcevic2016}
V.~Vukcevic, Numerical modelling of coupled potential and viscous flow for
  marine applications, Ph.D. thesis, University of Zagreb, Zagreb (2016).

\bibitem{Vukcevic2017}
V.~Vuk{\v c}evi{\'c}, H.~Jasak, I.~Gatin, Implementation of the {{Ghost Fluid
  Method}} for free surface flows in polyhedral {{Finite Volume}} framework,
  Comput. Fluids 153 (2017) 1--19.
\newblock \href {http://dx.doi.org/10.1016/j.compfluid.2017.05.003}
  {\path{doi:10.1016/j.compfluid.2017.05.003}}.

\bibitem{Huang2007}
J.~Huang, P.~M. Carrica, F.~Stern, Coupled ghost fluid/two-phase level set
  method for curvilinear body-fitted grids, Int. J. Numer. Methods Fluids
  55~(9) (2007) 867--897.
\newblock \href {http://dx.doi.org/10.1002/fld.1499}
  {\path{doi:10.1002/fld.1499}}.

\bibitem{Hoang2013}
D.~A. Hoang, V.~van Steijn, L.~M. Portela, M.~T. Kreutzer, C.~R. Kleijn,
  Benchmark numerical simulations of segmented two-phase flows in microchannels
  using the {{Volume}} of {{Fluid}} method, Comput. Fluids 86 (2013) 28--36.

\bibitem{Popinet2009}
S.~Popinet, An accurate adaptive solver for surface-tension-driven interfacial
  flows, J. Comput. Phys. 228~(16) (2009) 5838--5866.
\newblock \href {http://dx.doi.org/10.1016/j.jcp.2009.04.042}
  {\path{doi:10.1016/j.jcp.2009.04.042}}.

\bibitem{Francois2006}
M.~M. Francois, S.~J. Cummins, E.~D. Dendy, D.~B. Kothe, J.~M. Sicilian, M.~W.
  Williams, A balanced-force algorithm for continuous and sharp interfacial
  surface tension models within a volume tracking framework, J. Comput. Phys.
  213 (2006) 141--173.

\bibitem{Ivey2015}
C.~B. Ivey, P.~Moin, Accurate interface normal and curvature estimates on
  three-dimensional unstructured non-convex polyhedral meshes, J. Comput. Phys.
  300 (2015) 365--386.
\newblock \href {http://dx.doi.org/10.1016/j.jcp.2015.07.055}
  {\path{doi:10.1016/j.jcp.2015.07.055}}.

\bibitem{Albadawi2013}
A.~Albadawi, D.~Donoghue, A.~Robinson, D.~Murray, Y.~Delaur{\'e}, Influence of
  surface tension implementation in {{Volume}} of {{Fluid}} and coupled
  {{Volume}} of {{Fluid}} with {{Level Set}} methods for bubble growth and
  detachment, Int. J. Multiph. Flow 53 (2013) 11--28.
\newblock \href {http://dx.doi.org/10.1016/j.ijmultiphaseflow.2013.01.005}
  {\path{doi:10.1016/j.ijmultiphaseflow.2013.01.005}}.

\bibitem{Yamamoto2016}
T.~Yamamoto, Y.~Okano, S.~Dost, Validation of the {{S}}-{{CLSVOF}} method with
  the density-scaled balanced continuum surface force model in multiphase
  systems coupled with thermocapillary flows, Int. J. Numer. Methods Fluids
  83~(3) (2016) 223--244.
\newblock \href {http://dx.doi.org/10.1002/fld.4267}
  {\path{doi:10.1002/fld.4267}}.

\bibitem{Dianat2017}
M.~Dianat, M.~Skarysz, A.~Garmory, A {{Coupled Level Set}} and {{Volume}} of
  {{Fluid}} method for automotive exterior water management applications, Int.
  J. Multiph. Flow 91 (2017) 19--38.
\newblock \href {http://dx.doi.org/10.1016/j.ijmultiphaseflow.2017.01.008}
  {\path{doi:10.1016/j.ijmultiphaseflow.2017.01.008}}.

\bibitem{Haghshenas2017a}
M.~Haghshenas, J.~A. Wilson, R.~Kumar, Algebraic coupled level set-volume of
  fluid method for surface tension dominant two-phase flows, Int. J. Multiph.
  Flow 90 (2017) 13--28.
\newblock \href {http://dx.doi.org/10.1016/j.ijmultiphaseflow.2016.12.002}
  {\path{doi:10.1016/j.ijmultiphaseflow.2016.12.002}}.

\bibitem{Galusinski2008}
C.~Galusinski, P.~Vigneaux, On stability condition for bifluid flows with
  surface tension: {{Application}} to microfluidics, J. Comput. Phys. 227~(12)
  (2008) 6140--6164.
\newblock \href {http://dx.doi.org/10.1016/j.jcp.2008.02.023}
  {\path{doi:10.1016/j.jcp.2008.02.023}}.

\bibitem{Denner2015}
F.~Denner, B.~G. {van Wachem}, Numerical time-step restrictions as a result of
  capillary waves, J. Comput. Phys. 285 (2015) 24--40.
\newblock \href {http://dx.doi.org/10.1016/j.jcp.2015.01.021}
  {\path{doi:10.1016/j.jcp.2015.01.021}}.

\bibitem{Zinkle1998}
S.~J. Zinkle, Summary of {{Physical Properties}} for {{Lithium}},
  {{Pb}}-{{17Li}}, and ({{LiF}})n\textbullet{}{{BeF2 Coolants}}, in: {{APEX
  Study Meeting}}, 1998, pp. 1 -- 8.

\bibitem{Fazio2015}
C.~Fazio, V.~Sobolev, A.~Aerts, S.~Gavrilov, K.~Lambrinou, P.~Schuurmans,
  A.~Gessi, P.~Agostini, A.~Ciampichetti, L.~Martinelli, et~al., Handbook on
  lead-bismuth eutectic alloy and lead properties, materials compatibility,
  thermal-hydraulics and technologies., Tech. Rep. 7268, Organisation for
  Economic Co-Operation and Development (2015).

\bibitem{Raseman1960}
C.~J. Raseman, H.~Susskind, G.~Farber, W.~McNulty, F.~Salzano, Engineering
  experience at {{Brookhaven National Laboratory}} in handling fused chloride
  salts, Tech. Rep. BNL 627, {Brookhaven National Laboratory} (1960).

\bibitem{Janz1979a}
G.~J. Janz, C.~B. Allen, N.~P. Bansal, R.~M. Murphy, R.~P.~T. Tomkins, Physical
  Properties Data Compilations Relevant to Energy Storage. {{II}}. Molten
  Salts: {{Data}} on Single and Multi-Component Salt Systems, {U. S. Department
  of Commerce}, 1979.

\bibitem{Janz1988}
G.~J. Janz, Thermodynamic and Transport Properties for Molten Salts:
  Correlation Equations for Critically Evaluated Density, Surface Tension,
  Electrical Conductance, and Viscosity Data, no.~17 in J. Phys. Chem. Ref.
  Data, American Chemical Society and the American Institute of Physics, 1988.

\bibitem{Girifalco1957}
L.~A. Girifalco, R.~J. Good, A theory for the estimation of surface and
  interfacial energies. {{I}}. {{Derivation}} and application to interfacial
  tension, J. Phys. Chem. 61~(7) (1957) 904--909.

\bibitem{Good1970}
R.~J. Good, E.~Elbing, Generalization of theory for estimation of interfacial
  energies, Ind. Eng. Chem. 62~(3) (1970) 54--78.

\bibitem{Shaikhmahmud1953}
N.~Shaikhmahmud, Interfacial Tensions of Molten Metal-Molten Salt Systems :
  Bismuth against {{KCl}}-{{LiCl}} Eutectic Mixture / by {{N}}.{{S}}.
  {{Shaikhmahmud}}, {{C}}.{{F}}. {{Bonilla}}., {United States Atomic Energy
  Commission}, New York, 1953.

\bibitem{Roy1998}
R.~R. Roy, T.~A. Utigard, Interfacial tension between aluminum and
  {{NaCl}}-{{KCl}}-based salt systems, Metall. Mater. Trans. B 29~(4) (1998)
  821--827.

\bibitem{Roy1997}
R.~Roy, Y.~Sahai, Interfacial tension between {{Aluminium Alloy}} and {{Molten
  Salt Flux}}, Mater. Trans. 38~(6) (1997) 546--552.

\bibitem{Gasior2014}
W.~Gasior, Viscosity modeling of binary alloys: Comparative studies, Calphad 44
  (2014) 119--128.

\bibitem{Cornwell1971}
K.~Cornwell, The thermal conductivity of molten salts, J. Phys. Appl. Phys.
  4~(3) (1971) 441.

\end{thebibliography}

\clearpage
\appendix
\section{Spurious velocities}\label{ch:spurious}
Spurious currents are unphysical velocities well known in volume of
fluid (VOF) simulations \cite{Raeini2012, Harvie2006,Abadie2015}. One source of
such artificial flows is an imbalance between the pressure and density gradient
(eqn. \ref{eqn:nse}) due to the continuous interpolation of density
(or viscosity) \cite{Vukcevic2016,Vukcevic2017}. This type of
parasitic currents is strong for large density jumps and appears
predominantly in the lighter phase \cite{Huang2007}.
A second type of spurious currents is caused by the imbalance between
the surface tension force (Eq.~\ref{eqn:fst}) and pressure
gradient \cite{Hoang2013, Popinet2009} due to a poor calculation of
the interface curvature (Eq.~\ref{eqn:kappa}) \cite{Hoang2013,
Francois2006}. Moreover, this surface tension term may contain an
erroneous rotational component that cannot be balanced by the
irrotational pressure term \cite{Harvie2006}. This will be an
additional source of spurious currents. Finally, also the interface
compression term in
the phase transport equation \cite{Rusche2002} could lead to some
unphysical currents.

Naturally, all spurious currents can be reduced by lowering the
pressure residual. Type one spurious velocities may be eliminated by an
exact discretization of the pressure jump, i.e. by using the idea of the ghost
fluid method \cite{Huang2007,Vukcevic2017}. Interface tension related erroneous
flows are best treated by improving the curvature calculation. This
might be done using an (additional) height or level set function
\cite{Hoang2013,
Ivey2015,Albadawi2013,Yamamoto2016,Dianat2017,Haghshenas2017a}. Furthermore,
a posterior damping of 
spurious velocities is possible by smoothing the curvature field and
filtering interface-parallel velocities \cite{Raeini2012}.

Finally, the interface tension $\gamma$ is discretized
explicitly when using the CSF model \cite{Brackbill1992}. Therefore,
the capillary number must be respected as an additional time step
constraint \cite{Galusinski2008, Denner2015}:
\begin{linenomath*}
\begin{equation}
\Delta t = \sqrt{\frac{(\rho_A + \rho_B)\Delta
x^3}{2\pi\gamma_\text{max}}}\cdot\text{Co}_\text{cap}.
\end{equation}
\end{linenomath*}
Here, $\rho_A$ and $\rho_B$ denote the two smallest densities, $\Delta
x^3$ the cell volume, $\gamma_\text{max}$ the largest interface tension and
Co$_\text{cap}$ the capillary Courant number.

Implementing this time step restriction (for Co$_\text{cap}=0.5$),
removing the artificial interface compression and reducing the pressure
residual to $10^{-8}$ we were able to reduce spurious velocities from
\SI{1}{\centi\meter\per\second} to \SI{30}{\micro\meter\per\second} in
our setup.

\section{Constants for heat conduction model}\label{ch:constants_conduction}
\begin{align*}
c_1 &=\frac{(\dot{q}_\text{r}''+\dot{q}_\text{r}'''(h_\text{P}- h_\text{0})
-\dot{q}_{\Omega}'''h_\text{P})}{k_\text{P}}+\frac{k_\text{E}}{k_\text{P}}
c_\text{5} \\[2em]
c_2 &= T_\text{b}\\[2em]
c_3 &= \frac{(\dot{q}_\text{r}'''-\dot{q}_{\Omega}''')
h_\text{P}+\dot{q}_\text{r}''}{k_\text{P}}+\frac{k_\text{E}}{k_\text{P}}
c_\text{5} \\[2em]
c_4 &= T_\text{b} - \frac{\dot{q}_\text{r}'''h_\text{r}^2}{2k_\text{P}}\\[2em]
c_5 &=\frac{\dot{q}_{\Omega, \text{E}}'''(2k_\text{P} k_\text{E}  \left(  
h_\text{E}h_\text{N}- h_\text{E}^{2} \right) + k_\text{P} k_\text{N}  \left( 
h_\text{E}^{2}- h_\text{P}^{2}\right) )+ k_\text{E} k_\text{N} \left(
\dot{q}_\text{r}'''h_\text{r}^{2} - 2 \dot{q}_\text{r}'' h_\text{P} +
h_\text{P}^{2} \left(2 \dot{q}_{\Omega}''' -
\dot{q}_\text{r}'''\right)\right)}{2 k_\text{E} \left( h_\text{P} k_\text{E}
k_\text{N} - h_\text{P} k_\text{P} k_\text{N}- h_\text{E} k_\text{P} k_\text{E}
+ h_\text{E} k_\text{P} k_\text{N} + h_\text{N} k_\text{P} k_\text{E}\right)}
\\[2em]
c_6 &=T_\text{b}+h_\text{P}
c_5(\frac{k_\text{E}}{k_\text{P}}-1)+\frac{\dot{q}_\text{r}'''(h_\text{P}^2-h_\text{r}^2)+2h_\text{P}(\dot{q}_\text{r}''
-\dot{q}_{\Omega}'''h_\text{P}) }{2
k_\text{P}}+\frac{\dot{q}_{\Omega}'''h_\text{P}^2 }{2 k_\text{E}}\\[2em]
c_7
&=\frac{k_\text{E}}{k_\text{N}}(c_\text{5}-\frac{\dot{q}_{\Omega}'''h_\text{E}
}{k_\text{E}})  \\[2em]
c_8 &=T_\text{b} -h_\text{N} c_\text{7}  \\[2em]
\end{align*}

\section{Material properties and layer height}\label{ch:material_properties}
In this appendix the thermo-physical and transport properties are collected.\\
The material properties of \ce{Li} are taken from Zinkle et al. \cite{Zinkle1998},
and for Bi by Fazio et al. \cite{Fazio2015}.
Specific heat capacity and viscosity of the molten salt mixture (\ce{KCl-LiCl}) 
are evaluated with the formulation suggested by Raseman et al.
\cite{Raseman1960}, 
for the other properties the values given by 
Janz are used \cite{Janz1979a,Janz1988}.
All thermo-physical and transport properties of 
the pure materials of the cell are evaluated at \SI{450}{\celsius} 
and summarized in Tab.~\ref{tab:properties}. \\
The interface tensions can be approximated 
from the surface tensions of the pure substances, 
using  the rule proposed by  Girifalco and 
Good \cite{Girifalco1957,Good1970}:
\begin{linenomath*}
\begin{equation}
\gamma = \gamma_i+\gamma_j - 2\phi\sqrt{\gamma_i\gamma_j}.
\end{equation}
\end{linenomath*}
The interaction parameter $\phi$ for \ce{Bi}$|$\ce{KCl-LiCl} 
is given by Shaikhmahmud et al. 
\cite{Shaikhmahmud1953} as $\phi=0.58$. No information is available for the
couple \ce{Li}$|$\ce{KCl-LiCl}. However, the values for similar material combinations 
(e.g. 0.41 for Al$|$cryolite \cite{Roy1998}, 0.51 for Al$|$NaCl-KCl \cite{Roy1997}) suggest that 
assuming $\phi=0.5$ is reasonable. Using $\phi=0.5$ we obtain an interface
tension of $\gamma = \SI{0.196}{\newton.\meter^{-1}}$  for
Li$|$\ce{KCl-LiCl}
and $\gamma = \SI{0.275}{\newton.\meter^{-1}}$ for \ce{KCl-LiCl}$|$Bi. \\
The material properties of the \ce{Bi-Li} alloy in the concentration range of
interest are not available in literature. We assume a pure bismuth positive
electrode in full charge condition ($x_\text{Li} = \SI{1}{\percent}$). 
For higher concentration of \ce{Li} the values are approximated from the 
properties of the pure components. The mixture densities 
are calculated by the Vegard's law using the mole 
fraction \cite{Fazio2015}. These values are used to calculate the positive electrode
thickness. The specific heat capacity of 
the \ce{Bi-Li} alloy is calculated by Neumann-Kopp's law 
with the mass fraction. 
The surface tension of \ce{Bi-Li} alloy 
is assumed to be  
equal to the one of pure bismuth.
The viscosity of the mixture is calculated with a simple linear 
interpolation using the mole fractions, without taking into account 
the correction based on the thermodynamic potential proposed by 
Gasior et al. \cite{Gasior2014}. Regarding electrical and 
thermal conductivity, the values of pure bismuth are employed also 
for the alloy.
In Tab.~\ref{tab:charge_states} the dimensions of the cell 
at different charge states as well as the material properties of 
the mixture \ce{Bi-Li} are provided.

\begin{table}[h!]
\centering
\caption{Thermo-physical and transport properties of the pure substances at
$T=\SI{450}{\celsius}$
\cite{Zinkle1998,Janz1979a,Janz1988,Fazio2015,Raseman1960,Cornwell1971}.}\label{tab:properties}

\begin{tabular}{llrrr}
\hline
\multicolumn{1}{c}{property}                                    &
\multicolumn{1}{c}{unit}            & \multicolumn{1}{c}{Li} &
\multicolumn{1}{c}{LiCl-KCl} & \multicolumn{1}{c}{Bi} \\ \midrule
$\rho_\text{ref} $& \si{\kilogram.\meter^{-3}}         &                  491.3
&                         1648 &                   9843 \\
$\beta$                                                         &
\SI{e-4}{\per\kelvin}          &                   1.80 &                       
3.20 &                   1.24 \\
$c_p$                                                           &
\si{\joule.\kilogram^{-1}.\kelvin^{-1}} &                   4237 &              
1330 &                    136 \\
$\gamma$                                                        &
\si{\newton\meter^{-1}}                 &                  0.292 &                  
0.130 &                  0.362 \\  \midrule
$\nu$                                                           &
\SI{e-7}{\square\meter.\second^{-1}} &                   7.13 &                 
19.8 &                   1.33 \\
$\rho_\text{el}$                                                & 
\SI{e-6}{\ohm\meter}            &                  0.367 &                      
6358 &                   1.39 \\ 
$k$                                                             &
\si{\watt.\meter^{-1}.\kelvin^{-1}}                              &              
51.9 &                         0.69 &                   14.2 \\\hline
\end{tabular}\vspace{3pt}\end{table}

\begin{table}[h!]
\centering
\caption{Mole fractions of lithium in bismuth, layer heights and \ce{Bi-Li}
mixture properties (of the positive electrode) at different charge
states
\cite{Fazio2015,Zinkle1998}. } \label{tab:charge_states}
\begin{tabular}{llrrr}
\hline
\multicolumn{1}{c}{property} & \multicolumn{1}{c}{unit}             &          
\multicolumn{3}{c}{cases studied}            \\ \midrule
$x_\text{Li}$                &  \si{\percent}                                  
&     1 &                    10 &                   38 \\
$\Delta h_\text{P}$                        & \si{\milli\meter}                 
&    20 &                   21.3 &                   27.8 \\
$\Delta h_\text{E}$                        & \si{\milli\meter}                 
&     5 &                      5 &                      5 \\
$\Delta h_\text{N}$                        & \si{\milli\meter}                 
&    40 &                   38.5 &                   31.8
\\\midrule
$\rho_\text{ref}$                     &\si{\kilogram.\meter^{-3}}              
&  9843 &         9238 &         7222 \\
$\beta$                      & \SI{e-4}{\per\kelvin}                        & 
1.24 &         1.30 &         1.48 \\
$c_p$                        & \si{\joule.\kilo\gram^{-1}.\kelvin^{-1}} &   136
&        151.3 &        218.0 \\ 
$\gamma$                                                        &
\si{\newton\meter^{-1}}                 &         0.362           &      0.362      
&                  0.362 \\  \midrule
$\nu$                        &\SI{e-7}{\square\meter .\second^{-1}}            
&  1.33 &          1.31 &          1.31 \\
$\rho_\text{el}$             & \SI{e-6}{\ohm.\meter}               &  1.39 &
1.39 &   1.39 \\
$k$                          & \si{\watt.\meter^{-1}.\kelvin^{-1}}             
&  14.2 &   14.2 & 14.2 \\\hline
\end{tabular}\vspace{3pt}
\end{table}
\end{document}